\documentclass[manuscript,screen,acmsmall]{acmart}
\usepackage{tablefootnote}
\usepackage{stfloats}
\usepackage{multirow}
\usepackage{subfigure} 
\usepackage{makecell}
\usepackage{caption}
\usepackage{subfigure}
\usepackage{arydshln}
\usepackage{booktabs}
\usepackage{multirow}
\usepackage{tablefootnote}
\usepackage{setspace}
\usepackage{hyperref}
\usepackage{color}
\usepackage{xcolor}
\usepackage{lineno}

\definecolor{rewrite}{RGB}{148,0,211}
\AtBeginDocument{%
  \providecommand\BibTeX{{%
    \normalfont B\kern-0.5em{\scshape i\kern-0.25em b}\kern-0.8em\TeX}}}

\begin{document}

\title{NIR-Prompt: A Multi-task Generalized Neural Information Retrieval Training Framework}


\author{Shicheng Xu}
\affiliation{%
   \institution{Data Intelligence System Research Center, Institute of Computing Technology, CAS; University of Chinese Academy of Sciences}
  \city{Beijing}
  \country{China}}
\email{xschit@163.com}

\author{Liang Pang}
\authornote{Corresponding author}
\affiliation{%
   \institution{Data Intelligence System Research Center, Institute of Computing Technology, CAS}
  \city{Beijing}
  \country{China}}
\email{pangliang@ict.ac.cn}

\author{Huawei Shen}
\affiliation{%
   \institution{Data Intelligence System Research Center, Institute of Computing Technology, CAS; University of Chinese Academy of Sciences}
  \city{Beijing}
  \country{China}}
\email{shenhuawei@ict.ac.cn}

\author{Xueqi Cheng}
\affiliation{%
   \institution{CAS Key Lab of Network Data Science and Technology, Institute of Computing Technology, CAS; University of Chinese Academy of Sciences}
  \city{Beijing}
  \country{China}}
\email{cxq@ict.ac.cn}
\begin{abstract}
Information retrieval aims to find information that meets users' needs from the corpus. Different needs correspond to different IR tasks such as document retrieval, open-domain question answering, retrieval-based dialogue, etc., while they share the same schema to estimate the relationship between texts. It indicates that a good IR model can generalize to different tasks and domains. However, previous studies indicate that state-of-the-art neural information retrieval (NIR) models, e.g., pre-trained language models (PLMs) are hard to generalize. {It is mainly because} the end-to-end fine-tuning paradigm makes the model overemphasize task-specific signals and domain biases but loses the ability to capture generalized essential signals. To address this problem, we propose a novel NIR training framework named NIR-Prompt for retrieval and reranking stages based on the idea of decoupling signal capturing and combination. NIR-Prompt exploits Essential Matching Module (EMM) to capture the essential matching signals and gets the description of tasks by Matching Description Module (MDM). The description is used as task-adaptation information to combine the essential matching signals to adapt to different tasks. Experiments under in-domain multi-task, out-of-domain multi-task, and new task adaptation settings show that NIR-Prompt can improve the generalization of PLMs in NIR for both retrieval and reranking stages compared with baselines.

\end{abstract}
\newcommand\blfootnote[1]{%
\begingroup
\renewcommand\thefootnote{}\footnote{#1}%
\addtocounter{footnote}{-1}%
\endgroup
}
\blfootnote{This article is an extension of reference~\cite{match-prompt} on CIKM 2022. The previous conference version focuses on only the generalization ability of the reranking stage in information retrieval. 
However, the classic pipeline of IR can be divided into two stages including retrieval and reranking. Retrieval aims to efficiently and accurately filter candidate subset texts from the large-scale corpus. Reranking aims to rank the candidate subset texts more finely and return them to the users.
Compared with the previous work, (1) we first formulate our approach as a generalized neural information retrieval framework based on the idea of decoupling the process of signal capturing and signal combination, (2) then extending it to improve the generalization ability of the entire information retrieval pipeline, including retrieval and reranking, (3) experiment results on retrieval stage, reranking stage, and entire information retrieval pipelines on diverse datasets (eighteen public datasets and the heterogeneous benchmark) demonstrate the effectiveness of our framework in improving the generalization ability of neural information retrieval. }
\footrule



\maketitle

\section{Introduction} \label{intro}
Information retrieval (IR) is the fundamental task to find the target document from the large-scale resources to meet the users' information needs and has been applied to many downstream tasks such as document retrieval (DR)~\cite{adhoc}, open-domain question answering (QA)~\cite{open-domain-qa,adaptive_qa}, retrieval-based dialogue (RD)~\cite{ac}. {Traditional information retrieval methods such as BM25 exploit word-to-word exact matching to score the relevance between texts and rank them but these methods lack the capture of semantics. Recently, deep neural network has been introduced into information retrieval such as dense retrieval for retrieval~\cite{dpr} and cross-attention for reranking~\cite{monobert}. It can capture the semantic relationship between texts and achieves significant improvement when there is enough in-domain training data. However, previous studies have shown that the multi-task generalization ability of neural information retrieval (NIR) models is poor~\cite{beir,challenge_in_dr}, even worse than traditional word-to-word exact matching methods. }In general, most NIR models can only achieve good performance when trained and evaluated on a single specific dataset, while having poor generalization ability to different domains and tasks~\cite{linguistic,beir}.

{To address the aforementioned issue in neural information retrieval models, we begin by defining and evaluating multi-task generalization. In this study, the multi-task generalization of a NIR model is categorized into three levels: 1) In-domain multi-task: The model has access to various datasets encompassing multiple IR tasks such as QA, RD, and DR. Evaluation is performed on each provided dataset. This level assesses the model's capability to learn from diverse IR tasks and apply that knowledge effectively within the respective domain of each task. 2) Out-of-domain multi-task: The model can access the same datasets as in the in-domain multi-task level. However, it is evaluated on unseen datasets from different domains. This level examines the model's ability to transfer learned knowledge to new domains, for instance, transitioning from financial QA to medical QA. 3) New task adaptation: This level involves making a task invisible to the model during training, such as QA, while training the model on other tasks like RD and DR. The model is provided with few-shot examples from the unseen task to adapt and generalize to this new task. This level particularly focuses on assessing the model's capability for new task generalization, which is the most challenging level for IR models.}

\begin{figure}[htbp]
  \centering
  \includegraphics[width=0.8\linewidth]{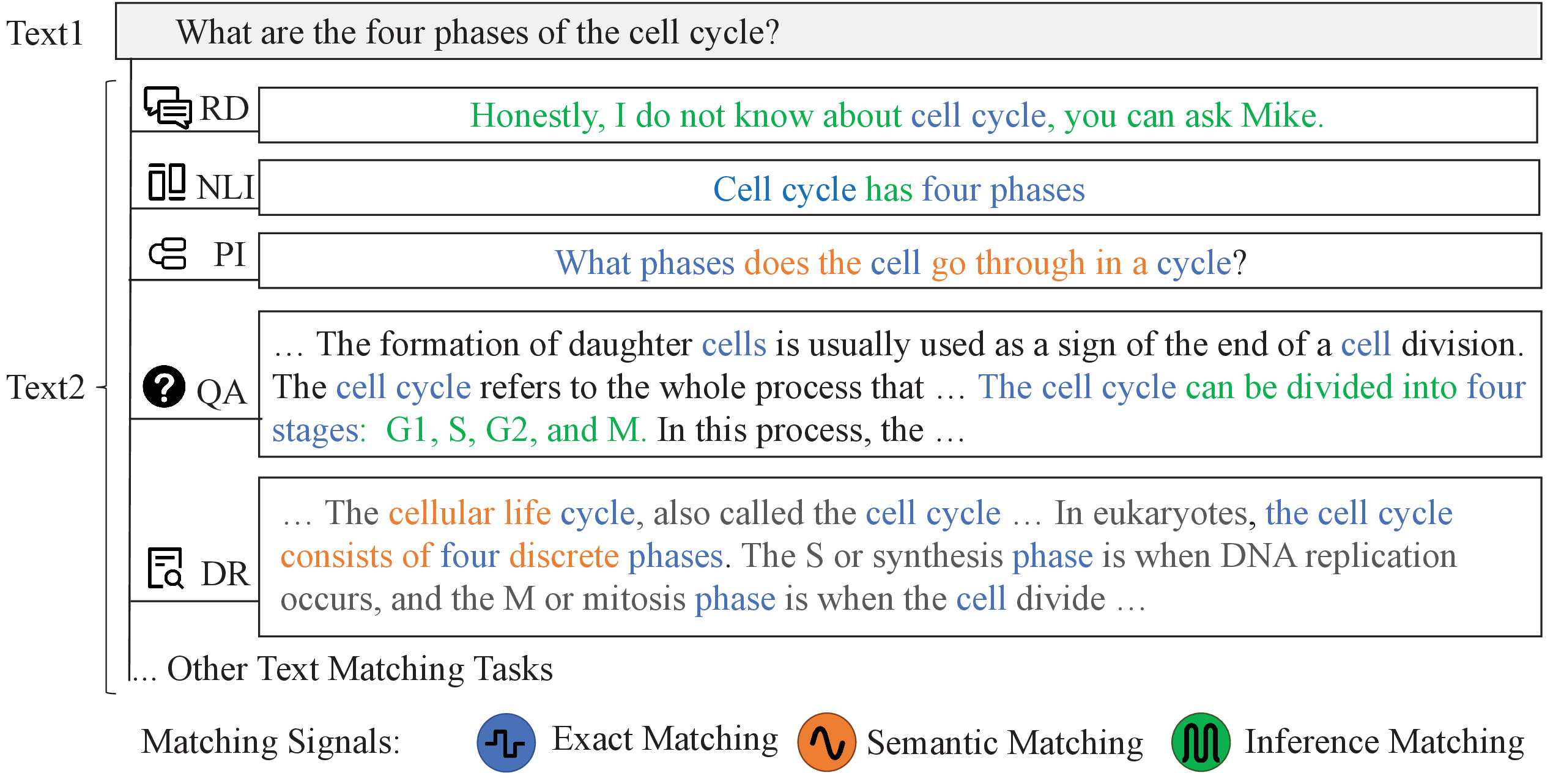}
  \caption{Matching texts and signals for {\itshape "What are the four phases of the cell cycle?"} in different tasks.}
  \label{figure:example}
\end{figure}

{Then, we point out that the core sub-task of the state-of-the-art (i.e. PLM-based) NIR models is text matching. Text matching aims to estimate the relevance score between two texts for a specific task. Generalization in neural information retrieval can be converted to the generalization of text matching. The foundation for the NIR models to estimate the relevance score between texts in text matching is the matching signal. Although there are certain differences in the matching signals for different retrieval tasks, there are matching signals that are shared across tasks and we call them essential matching signals. Since paraphrase identification (PI)~\cite{pi} and natural language inference (NLI)~\cite{nli} are also two mainstream text matching tasks (but are not IR tasks) and can also provide the shared essential matching signals, we also incorporate them into the analysis of matching signals. We take three most common and unambiguous essential matching signals across five text matching tasks as examples, namely exact matching signals, semantic matching signals, and inference matching signals. As shown in Figure~\ref{figure:example}, exact matching signals play a crucial role in all text matching tasks by focusing on word overlap. They are important in various text matching tasks, such as query term importance in DR\cite{drmm, deeprank}, lexical overlap features in PI\cite{pipqr}, and bigram counts in QA~\cite{open-domain-qa}. Semantic matching signals measure the semantic association between texts beyond just word overlap. Many studies prove that introducing semantic information into DR~\cite{dssm} and PI~\cite{matchpyramid} can improve the performance, especially for QA that dense retrieval~\cite{dense-retrival} is better than lexical retrieval by training on sufficient in-domain data, e.g. DPR~\cite{dpr} and ORQA~\cite{orqa}. Inference matching signals infer implication relationship. NLI, QA, and RD have to infer the implication relation between the two texts. For example, RD needs to confirm the reply is a logical consequence of the post and dialogue history, and QA needs to pick out the answer that the question hardly contains. From the applying perspective, the difference between text matching tasks is the different fusion and utilization of these matching signals. However, the current mainstream end-to-end fine-tuning paradigm makes the model overemphasize the task-specific matching signals and domain biases but loses the ability to capture the essential matching signals that can be used across different matching tasks and domains for IR, which reduces generalization to different tasks and domains.} Thus, if one matching model can capture the essential matching signals shared across tasks and combine them according to a specific matching task, its multi-task generalization ability will be improved. However, the main challenges to obtaining a good generalized text matching model lay in two folds, 1) how to capture the essential matching signals? 2) how to exploit these signals to apply the model to different matching tasks?

{Typical IR tasks mainly involve two stages, the first is \textbf{retrieval} that aims to recall the subset of candidate documents from the huge amount of resources, and the second is \textbf{reranking} that aims to further rank the retrieved subset more finely. In this paper, we propose a generalized NIR model training framework called NIR-prompt for both retrieval and reranking in IR. NIR-prompt captures and exploits essential matching signals based on the idea of decoupling the process of signal capturing and signal combination via prompt learning. }Specifically, NIR-Prompt transforms text matching tasks into the form of [MASK] prediction by constructing the prompt template added to input texts, which is more in line with the pre-training tasks of PLMs. For retrieval, the prompt tokens instruct the PLM to output token embedding at [MASK] to map queries and documents to the latent semantic space and then calculate the relevance score based on embedding similarities. For reranking, the two texts are concatenated with the constructed prompt template, and the relevance score between the texts is estimated by predicting the probability distribution of the output word at [MASK]. NIR-Prompt consists of an \textbf{Essential Matching Module (EMM)} and a \textbf{Matching Description Module (MDM)}. MDM maps the description of different matching tasks to a few prompt tokens through prompt engineering. EMM is trained on mixed datasets consisting of different text matching tasks, and the prompt tokens from the MDM are used as the task-adaptation tokens to guide the learning of essential matching signals in EMM to adapt them to different tasks. Diversifying text matching tasks help the model explore the essential matching signals instead of overemphasizing the data sample bias and task-specific signals to capture the shared information from tasks and address the first challenge. The prompt tokens of each task are added to the corresponding tasks to distinguish different tasks and adapt essential matching signals to the specific tasks to answer the second challenge. Besides, we design a simple but effective method to construct discriminative tokens for new tasks by combining the learned prompt tokens, which to some extent indicate the correlation between different matching tasks.

Experimental results on eighteen public datasets and BEIR (the heterogeneous benchmark for testing the generalization ability of retrieval models)~\cite{beir} show that our method yields better in-domain multi-task, out-of-domain multi-task, and new task adaptation performance for the retrieval stage, reranking stage, and entire information retrieval pipeline compared to the traditional fine-tuning paradigm. The results also indicate that NIR-Prompt has a stronger ability to distinguish tasks and utilize essential matching signals shared by multiple tasks. Both of them are beneficial for improving the multi-task generalization ability. 

To sum up, our contributions are as follows:
\begin{itemize}
\item{} {We propose that various text matching tasks have shared matching signals that can be used across different matching tasks and domains for information retrieval. These signals are essential for generalization in neural information retrieval models.}
\item{} {We propose a novel framework named NIR-prompt to implicitly capture and combine the essential matching signals to improve the generalization ability of NIR models for the retrieval stage, reranking stage, and the entire information retrieval pipeline. }
\item{} We collect eighteen datasets from diverse text matching tasks, which can serve as a benchmark for the multi-task generalization ability of NIR models. We evaluate our method on these datasets in three settings including in-domain multi-task, out-of-domain multi-task, and new task adaptation. Besides, we also test the zero-shot ability of our method on BEIR to further demonstrate the positive effect of our method on the generalization ability of NIR models. Code and datasets will be released at \url{https://github.com/xsc1234/NIR-Prompt/tree/main/}.
\end{itemize}

\section{PRELIMINARIES ABOUT NEURAL INFORMATION RETRIEVAL}
The mainstream neural information retrieval pipeline includes two stages: retrieval and reranking. In neural retrieval, dense retrieval is the most commonly used method that considers both efficiency and retrieval performance, so in this paper, we focus on dense retrieval for the retrieval stage. Neural reranking aims to model the relevance between texts with more fine-grained interaction function, and the most commonly used methods are generally based on the cross encoder, such as Transformer~\cite{attention}, so in this paper, we focus on the interaction-based models for the reranking stage. Details about them will be introduced as follows.

\subsection{Dense Retrieval}
Dense retrieval is the first stage in the information retrieval system that efficiently and accurately obtains candidate subsets from the massive document base.
Dense retrieval model is a dual-encoder structure that encodes the query and document into dense embeddings and estimates the relevance score by measuring the similarity between the two embeddings. For a text pair $(q,d)$, the relevance score $r$ can be computed by:
\begin{equation}
r = s(E_{q}(q),E_{d}(d)),
\nonumber
\end{equation}
where $E_q$ and $E_d$ are the encoders for $q$ and $d$ respectively, and $s$ is the similarity function such as inner-product and Euclidean distance. Based on this, the approximate nearest neighbor search algorithm (ANN-search~\cite{ann}) for embeddings is used to search from the document base efficiently.

\subsection{Reranking}
Reranking is the finer ranking stage for the candidate subsets retrieved from the first stage, which models the interaction between texts more complexly. Recently, with pre-trained language models (PLM) applied to various natural language processing tasks, the mainstream structure for reranking becomes the PLM-based model with concatenated text pairs. For a text pair $(q,d)$ the relevance score $r$ can be computed by:
\begin{equation}
r = Scoring(f(q \circ d)),
\nonumber
\end{equation}
where $\circ$ is the concatenation operator, $f$ is the interactive function such as self-attention~\cite{attention}, $Scoring$ estimated relevance score or category according to the interaction of the two texts.

\section{NIR-Prompt}
In this section, we describe the core idea of NIR-Prompt and the technical details of using it to build multi-task generalized neural information retrieval pipeline, including dense retrieval (Retriever-Prompt) and reranking (Reranker-Prompt).

\begin{figure}[t]
  \centering
  \includegraphics[width=\linewidth]{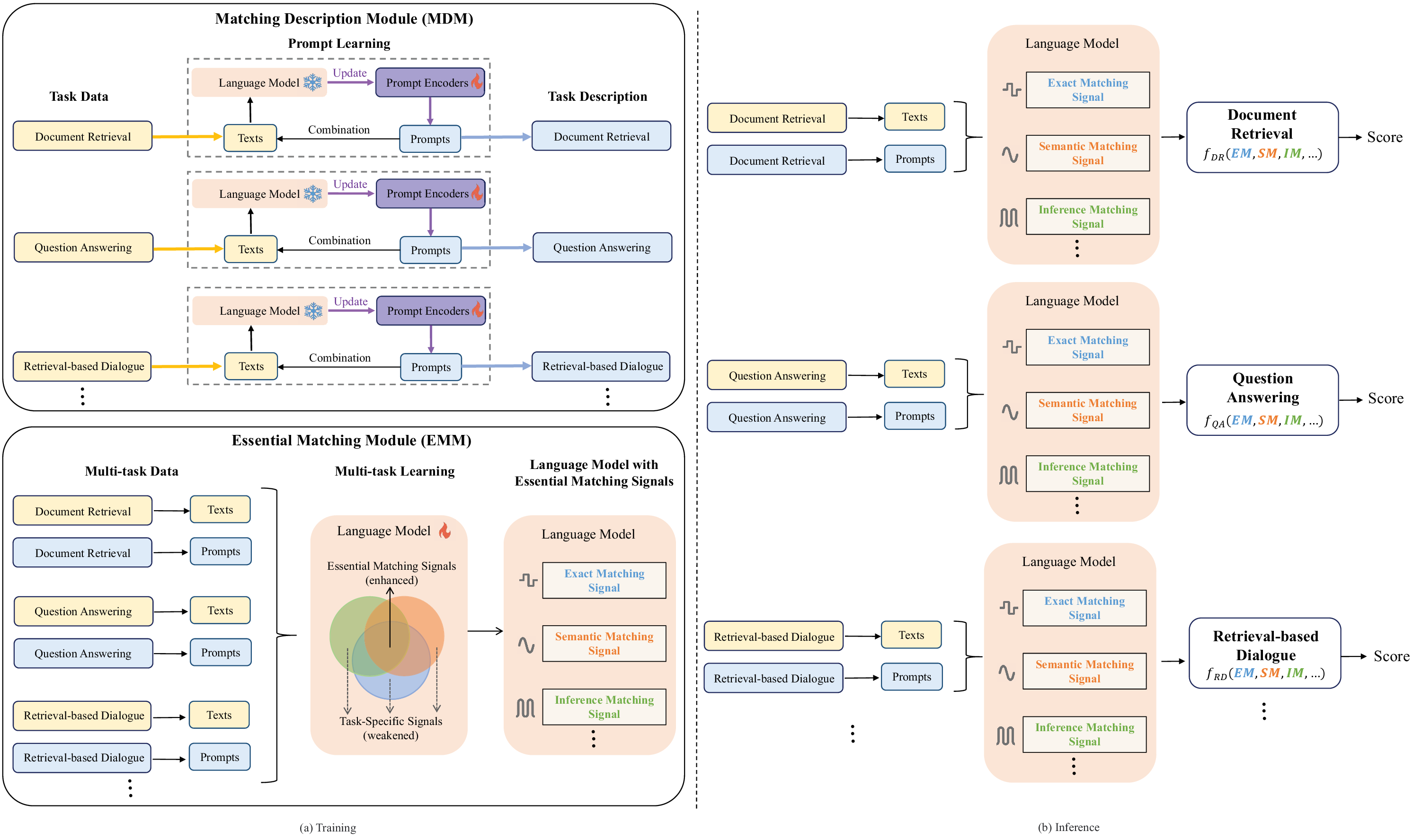}
  \caption{{The basic idea of NIR-Prompt. The Matching Description Module (MDM) exploits prompt learning to map the characteristics of each task to prompt tokens and use them as the task description. The Essential Matching Module (EMM) combines the description and the data of each task and mixes various tasks into mixed datasets. PLM performs multi-task learning on the mixed datasets to implicitly capture the essential matching signals shared across tasks and adapt them to the description of each task. Task description combines the essential matching signals to generalize to multiple tasks.}}
  \label{figure:motivation}
\end{figure}

\subsection{Basic Idea}
The basic idea of NIR-Prompt is shown in Figure~\ref{figure:motivation}. NIR-Prompt consists of an Essential Matching Module (EMM) and a Matching Description Module (MDM). EMM captures common and essential matching signals for various text matching tasks in information retrieval, such as exact matching signals, semantic matching signals, and inference matching signals. MDM obtains the descriptions of different tasks in PLM and uses the descriptions to guide the learning and combination of essential matching signals in EMM to adapt to different text matching tasks and domains. Specifically, in MDM, the descriptions of each task are mapped to a few prompt tokens by prompt engineering. These tokens contain the information of each task in PLM and can be used as the differentiating marks of each task to guide the model to adapt to different tasks in multi-task learning. In EMM, prompt tokens obtained in MDM are added to the input text of the corresponding specific task and PLM is trained on mixed datasets consisting of different tasks. High-diverse matching tasks prevent the model from fitting the data sample bias on a specific task so that the model can focus on learning the common and essential matching signals that can be used across domains and tasks. Besides, prompt tokens obtained in MDM help the multi-task model better combine essential matching signals to adapt to different tasks, both of them are beneficial for improving its multi-task generalization ability.
\subsection{Overall Framework}
\begin{figure}[t]
  \centering
  \includegraphics[width=0.95\linewidth]{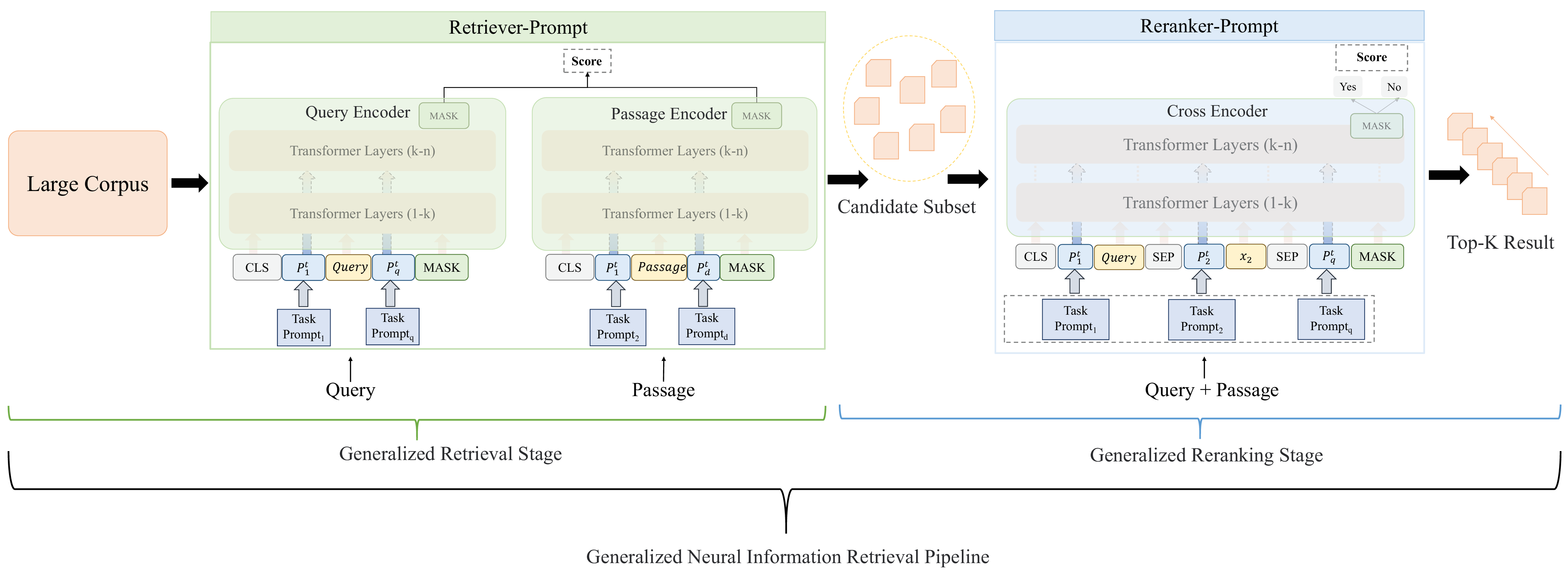}
  \caption{{Neural information retrieval pipeline of NIR-prompt. NIR-prompt consists of two stages, one is Retriever-Prompt for retrieval the other is Reranker-Prompt for reranking. In each stage, task prompts are used as the task description to combine the essential matching signals and generalize to multiple tasks.}}
  \label{figure:pipeline}
\end{figure}
Neural information retrieval pipeline consists of retrieval and reranking stages. In neural retrieval, dense retrieval is the most commonly used method that considers both efficiency and retrieval performance, so in this paper, we focus on dense retrieval for the retrieval stage. Neural reranking aims to model the relevance between texts with more fine-grained interaction function, and the most commonly used methods are generally based on the cross encoder, such as Transformer~\cite{attention}, so in this paper, we focus on the interaction-based models for the reranking stage. NIR-Prompt framework can be used to train the neural information retrieval pipeline (dense retrieval and reranking) based on the prompt learning paradigm so as to achieve multi-task generalization of the entire IR system. Our method in the retrieval stage is called Retriever-Prompt and in the reranking stage is called Reranker-Prompt. The pipeline of NIR-prompt is shown in Figure~\ref{figure:pipeline} and the relevant technical details will be introduced below. {We will introduce the Matching Description Module and Essential Matching Module in Retriever-Prompt and Reranker-Prompt respectively.}

\subsection{{Matching Description Module}}
\begin{figure}[t]
  \centering
  \includegraphics[width=\linewidth]{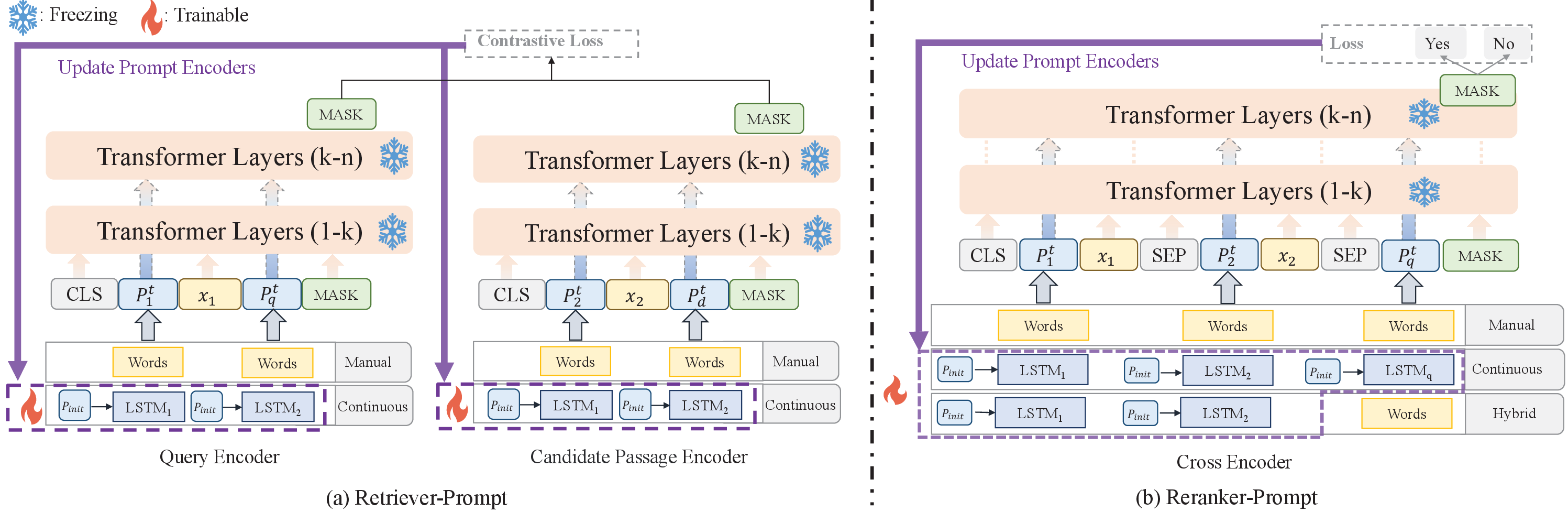}
  \caption{{Matching description module (MDM) in retrieval (a) and reranking (b) stage. Retriever-Prompt optimizes prompt encoders to generate prompt tokens ($P_1^{t}$, $P_2^{t}$, $P_q^{t}$ and $P_d^{t}$) to control transformer to map the text to dense embedding (token of [MASK]) with contrastive loss. Reranker-Prompt optimizes prompt encoders to  generate prompt tokens ($P_1^{t}$, $P_2^{t}$ and $P_q^{t}$) to control transformer to optimize the word probability distribution of the token of [MASK].}}
  \label{figure:mdm}
\end{figure}
{Matching description module (MDM) is used to generate the description (i.e. prompt tokens) for each task that reflects the knowledge of the PLM, which is used to guide the combination of the essential matching signals in the essential matching module to generalize to different tasks. The architecture of the matching description module for dense retrieval (a) and reranking (b) is shown in Figure~\ref{figure:mdm}, which consists of a frozen PLM and trainable prompt encoders. Prompt encoders map the description knowledge of the specific matching task in PLM into prompt tokens and update the parameters to generate the optimal tokens in training. Different tasks 
have different prompt tokens and are trained separately. These tokens play an important role in the forward computation of transformer layers, reflecting the descriptive knowledge of different tasks in PLM~\cite{prefix}, which can help the model describe different tasks and guide the learning and combination of essential matching signals in the essential matching module. To facilitate the introduction of these methods, we give the formal template of the input texts below. }

{In Retriever-Prompt, for a pair of input texts $(x_1,x_2) $, the template of the texts is:}
\begin{equation}
\begin{aligned}
x_{1}' = [\text{CLS}]\,P_1^{t}\,x_1\,P_q^{t}\,[\text{MASK}_1], \\
x_{2}' = [\text{CLS}]\,P_2^{t}\,x_2\,P_d^{t}\,[\text{MASK}_2].
\end{aligned}
\label{template}
\end{equation}
{The prompt tokens corresponding to each task $t$ are $P_1^{t}$, $P_q^{t}$, $P_2^{t}$ and $P_{d}^{t}$ for the input texts $x_{1}$ and $x_{2}$ respectively. They are the descriptions of the task. Specifically, $P_1^{t}$ ($P_2^{t}$) define the properties of $x_1$ ($x_2$) such as the query, passage, question, etc. $P_q^{t}$ ($P_d^{t}$) prompts PLM to generate the suitable representation for $x_1$ ($x_2$) by optimizing the embedding of the token at [MASK]. There is a huge gap between the dual encoder architecture in retrieval and the traditional prompt learning methods. It is meanly because that dual encoder architecture requires the two texts to be encoded independently, while traditional prompt learning methods need to input two texts to a pretrained language model (PLM) jointly and prompt the PLM to give the answer (cross encoder architecture). To solve this challenge, the training of MDM in Retriever-Prompt is updating the parameter of prompt encoders to generate these prompt tokens to optimize the embedding of the token at [MASK] exploiting the knowledge from the PLM. The embedding of the token at [MASK] is used as the text representation in dense retrieval and the optimization objective can be the function commonly used in dense retrieval, such as contrastive loss~\cite{contrastive}. The training details will be introduced in the following. }

{In Reranker-Prompt, different from retrieval, the reranking in IR is an interaction-based process that captures the matching relationship between texts in a more fine-grained way. Some text-pair classification tasks with more complex textual interaction information based on cross encoder such as Paraphrase Identification and Natural Language Inference are also considered in reranking to introduce more shared matching signals. Given a pair of input texts ($x_1$,$x_2$), we construct the input template to prompt PLM to judge the matching relationship between the input texts by predicting the word at the token of [MASK]. Specifically, the template for ($x_1$,$x_2$) for reranking is:}
\begin{equation}
x_{input} = [\text{CLS}]\,P_1^{t}\,x_1\,[SEP]\,P_2^{t}\,x_2\,[SEP]\,P_q^{t}\,[\text{MASK}],\label{x_input} \nonumber
 \end{equation}
{where $P_1^{t}$, $P_2^{t}$ and $P_q^{t}$ are the prompt tokens. $P_1^{t}$ and $P_2^{t}$ define the properties of $x_1$ and $x_2$ respectively, $P_q^{t}$ prompts PLM to judge the matching relationship of text pairs and fill the result into [MASK]. For the classification task, a mapper is used to map the class label to the output word at [MASK]. For the relevance score estimation task, the probability of the output word at [MASK] is used as the relevance score between $x_1$ and $x_2$. }

As for the methods of obtaining the value of prompt tokens in matching description module, we design three methods including manual prompt, continuous prompt, and hybrid prompt (specifically for reranking).

\subsubsection{Manual Prompt} \label{mp}

{Taking inspiration from LAMA~\cite{lama}, which utilizes manually crafted cloze templates to investigate knowledge in PLMs, the concept of manual prompts has been introduced for various NLP tasks like text classification~\cite{pet} and text generation~\cite{ge-pet}. Manual prompts exclusively consist of natural language. In Retriever-Prompt, specific manual prompts have been devised for different text matching tasks, as shown in Table~\ref{table:manual-prompt retrieval}. The prompts $P_1^{t}$ and $P_2^{t}$ are exploited to inform the PLM about the type of $x_1$ and $x_2$, respectively. $P_q^{t}$ and $P_d^{t}$ are the sentences that ask questions about the representation of the text according to the specific task. The representation of the texts $x_1$ and $x_2$ is the embedding of tokens $[MASK_1]$ and $[MASK_2]$. In Reranker-Prompt, on the other hand, showcases distinct manual prompts for different text matching tasks in Table~\ref{table:manual-prompt reranking}. $P_1^{t}$ and $P_2^{t}$ describe the attributes of the two texts individually, while $P_q^{t}$ poses questions pertaining to the relationship between the two texts, varying depending on the task. The anticipated answer is expected to be outputted at [MASK], possibly containing words like "yes" or "no". In comparison to continuous prompts, this approach employs natural language as prompts without necessitating additional training on prompt tokens, albeit at the expense of relying on human expertise, resulting in sub-optimal performance.}

\begin{table*}[t]
  \caption{Manual prompts in Retriever-Prompt. Handcrafted prompts are described in natural language.}
  \label{table:manual-prompt retrieval}
    \scalebox{0.8}{
  \begin{tabular}{llll}
    \toprule
    Task & $P_1^{t}$ & $P_2^{t}$ & $P_q^{t}$, $P_d^{t}$\\
    \midrule
   Document Retrieval & The query: & The passage: & Representation for document retrieval is:\\
    Question Answering & The question: & The passage: & Representation for question answering is:\\
    Retrieval-based Dialogue & The first sentence: & The second sentence: & Representation for retrieval-based dialogue is:\\

  \bottomrule
\end{tabular}}
\end{table*}

\begin{table*}[t]
  \caption{Manual prompts in Reranker-Prompt. Handcrafted prompts are described in natural language.}
  \label{table:manual-prompt reranking}
     \scalebox{0.77}{
  \begin{tabular}{llll}
    \toprule
    Task & $P_1^{t}$ & $P_2^{t}$ & $P_q^{t}$\\
    \midrule
   Document Retrieval & Query: & Passage: & Dose the passage include the content that matches the query?\\
    Question and Answer & Question: & Passage: & Does the passage include the answer of the question?\\
    Retrieval-based Dialogue & The first text: & The second text: & Can the second text reply to the first text?\\
    Paraphrase Identification & The first text: & The second text: & Do these two texts mean the same thing?\\
    Natural Language Inference & Premise: & Hypothesis: & Can the hypothesis be concluded from the premise?\\
  \bottomrule
\end{tabular}}
\end{table*}

\subsubsection{Continuous Prompt} \label{cp retrieval}

{Building upon the principles of P-tuning~\cite{p-tunning}, we present a novel approach for the optimization of prompt tokens in a continuous space. In our method, prompts are represented as trainable continuous vectors. Unlike P-tuning, our focus lies in enhancing the model's multi-task generalization capabilities specifically in text matching scenarios. The objective of acquiring prompt tokens for each task is not primarily to enhance prompt learning performance, but rather to let the prompt tokens give better descriptions of each specific text matching task. These descriptions are used to distinguish different tasks and adapt the essential matching signals to suit the specific requirements of each task. Consequently, in Retriever-Prompt, different from P-tuning, we introduce several improvements:}

\textbf{Prompt encoder}: {After pre-training, the word embedding in PLM has been very discrete. If the trainable prompt tokens are directly initialized with random distribution, and then optimized with stochastic gradient descent (SGD), it can only update parameters in a small neighborhood and is easy to fall into local optimum but cannot reach the global optimal point~\cite{p-tunning, convergence}. To address this limitation, a bidirectional Long Short-Term Memory (LSTM) network is employed in P-tuning to optimize all prompt token embeddings in a fully continuous space~\cite{p-tunning}. Different from it, our approach utilizes separate LSTMs for $P_1^{t}$ and $P_q^{t}$ ($P_2^{t}$ and $P_d^{t}$) respectively (as shown in Figure~\ref{figure:mdm}). The rationale behind this choice stems from the assumption made by LSTMs that there exists a certain dependency and sequential relationship among the prompt tokens. In other words, in a bidirectional LSTM, the tokens within $P_1^{t}$ and $P_q^{t}$ ($P_2^{t}$ and $P_d^{t}$) are interdependent. However, this interdependency, while present, unnecessarily restricts the representation of the embeddings of prompt tokens. It is because (1) The tokens of $P_1^{t}$ and $P_q^{t}$ can only be jointly generated by a single LSTM, and the number of trainable parameters is limited.
(2)	$P_q^{t}$ must obey the sequential relationship with $P_1^{t}$, which limits the value range of the embedding of tokens in $P_q^{t}$. Therefore, the representation of continuous prompt is limited by a single LSTM. By employing separate LSTMs for $P_1^{t}$ and $P_q^{t}$ ($P_2^{t}$ and $P_d^{t}$), we mitigate the interdependency between these three prompts, thus reducing the restriction on representation.}

\textbf{Fix the prompt tokens}: In P-tuning and many other prompt learning methods, prompt tokens are just added to input text and are alterable like other tokens in self-attention. Different from this, we let the embedding of trainable prompt tokens remain fixed in layers $1$ to $k$ in the self-attention. In these layers, prompt tokens can affect the calculation of PLM on input text, but the input text cannot affect prompt tokens. In the layers $k$ to $n$, prompt tokens are alterable. It is because if we do not impose any restrictions on prompt tokens, the embedding of the prompt tokens is affected by other input words and updates in every layer. This is not conducive to obtaining an abstract description of the specific text matching task, because there is no guarantee that the embeddings of prompt tokens are only determined by the prompt encoder due to the influence of other input words, which may cause the embedding to fit the data instead of the task. On the other hand, it is inconsistent with the mask prediction task in the pre-training process of PLM if we fix prompt tokens in each layer, which may lead to sub-optimal results when using the language model to optimize the token at [MASK]. Therefore, an intermediate layer needs to be determined to balance the two to achieve optimal performance. Based on our experiment, we find that the $11$-$th$ layer is an optimal boundary layer. We balance the description of the task and the interaction with other words in the text. In this way, prompt tokens can achieve better control of PLM and describe the task more comprehensively. 

The parameters of prompt encoders are updated through back-propagation thereby adjusting the embedding of the prompt tokens. Let $P_{init}$ be a randomly initialized vector used as the input to prompt encoder. $MLP_1^{t}(LSTM_1)$, $MLP_2^{t}(LSTM_2)$, $MLP_q^{t}(LSTM_q)$ and $MLP_d^{t}(LSTM_d)$ are the encoders of the corresponding prompt. Each encoder consists of a bidirectional LSTM and multilayer perceptron. These four encoders are optimized to obtain the continuous prompts respectively. The embeddings of $P_1^{t}$, $P_q^{t}$, $P_2^{t}$ and $P_d^{t}$ are obtained by:
\begin{equation}
    \begin{aligned}
      & P_1^{t} = MLP_1(LSTM_1(P_{init})),  \\
      & P_q^{t} = MLP_1(LSTM_1(P_{init})), \\
      & P_2^{t} = MLP_2(LSTM_2(P_{init})), \\
      & P_d^{t} = MLP_d(LSTM_d(P_{init})). \\
    \end{aligned}
    \label{prompt encoder} 
\end{equation}
The objective function of the prompt encoder is the in-batch contrastive loss and the details of training will be introduced in Section~\ref{training}.

In Reranker-Prompt, same as the continuous prompt for retrieval, LSTM is used to optimize the $P_1^{t}$, $P_2^{t}$ and $P_q^{t}$ respectively and the prompt tokens in first $k$ layer are fixed during the calculation of PLM. The most notable difference compared to retrieval is that the token of [MASK] is not used to represent the text but to output the word such as "yes" or "no" to determine the relationship between two texts and the relevance score can be obtained from the probability distribution of vocabulary. The details of training will be introduced in Section~\ref{training}. 

\subsubsection{Hybrid Prompt for Reranking}

We design the hybrid prompt for reranking specifically. In hybrid prompt, for $P_q^{t}$, instead of using a prompt encoder to generate a continuous vector, we express it as a natural language: "Do these two sentences match?". It is because, in the template of continuous prompt, text matching is converted into predicting the word at [MASK]. If the result is “yes”, these two texts match, otherwise they do not. In order to predict the word, the parameters of the three prompt encoders for $P_1^{t}$, $P_2^{t}$, and $P_q^{t}$ need to be updated. However, there is no prior information about the task other than training data. This leads to that although the prompt encoder can minimize loss during training, the task it describes may not be text matching. It just makes prompts control PLMs to extract features from the input texts and output the corresponding "yes" or "no", which is not conducive to generalizing to other datasets. Expressing $P_q^{t}$ as task-relevant natural language can control the task closer to text matching and facilitate $P_1^{t}$ and $P_2^{t}$ to describe different matching tasks, which is beneficial for improving generalization ability. The template contains both manual prompt ($P_q^{t}$) and continuous prompt ($P_1^{t}, P_2^{t}$). We call this method hybrid prompt. 

\subsection{Essential Matching Module}

\begin{figure}[t]
  \centering
  \includegraphics[width=\linewidth]{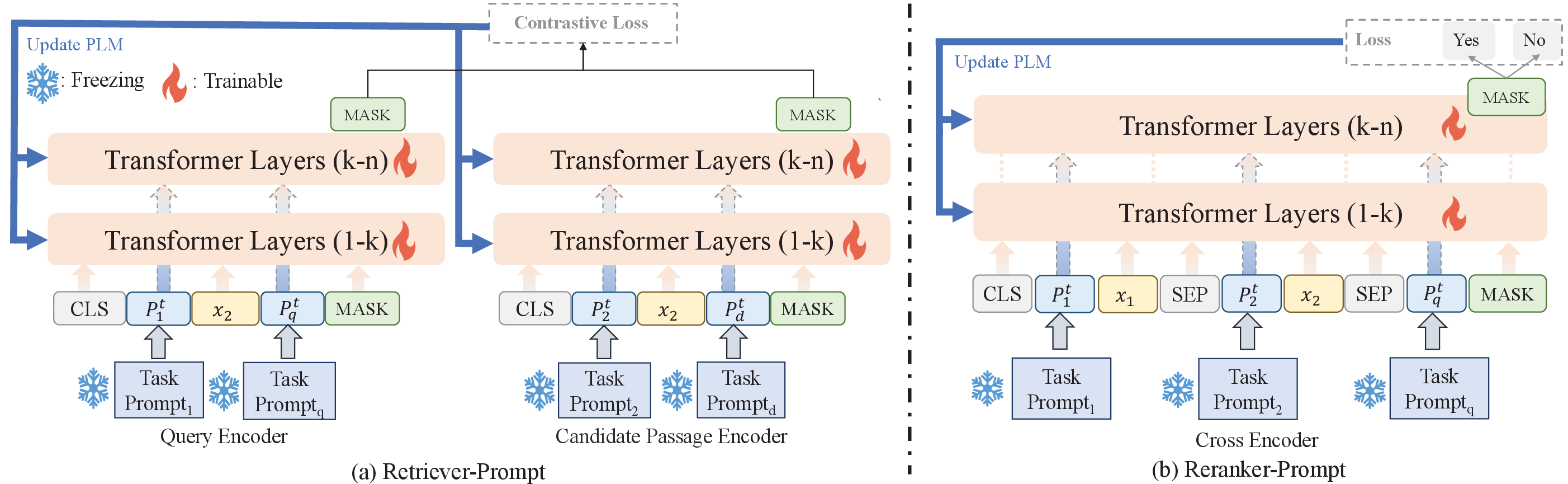}
  \caption{{Essential matching module (EMM) in retrieval and reranking stage. Task prompts are obtained from the prompt tokens of matching description module. EMM Optimizes transformer to capture the essential matching signals and adapt to different tasks under the control of prompt tokens.}}
  \label{figure:emm}
\end{figure}
{Essential matching module is used to capture essential matching signals across tasks and combine them with task descriptions obtained in MDM to adapt to different tasks via multi-task learning on the mixed datasets consisting of multiple tasks. In this paper, we propose that various IR tasks have their shared matching signals such as exact matching, semantic matching, and inference matching, etc. and we call them essential matching signals. The traditional end-to-end fine-tuning paradigm makes the model overemphasize the task-specific signals and domain biases but loses the ability to capture essential matching signals that can be used across different matching tasks and domains for IR. The architecture of the matching description module for dense retrieval (a) and reranking (b) is shown in Figure~\ref{figure:emm}, which consists of the trainable PLM and the task prompt tokens obtained from the matching description module. Multi-task learning can capture the information shared across tasks~\cite{mt-dnn}. In EMM, the PLM is trained on mixed datasets consisting of multiple high-diversity tasks to avoid overfitting the domain and task bias to capture the essential matching signals. During the training, task prompt tokens obtained from MDM contain the description of different tasks and guide the learning and combining of essential matching signals to adapt to different tasks. In the inference for different tasks, the task prompt tokens can be added to the input texts to control EMM to perform the specific task.}

\subsection{Training} \label{training}
Because of the different structures of the dual encoder in dense retrieval and cross encoder in reranking, we use in-batch contrastive loss and cross-entropy as loss functions for the training of Retriever-Prompt and Reranker-Prompt respectively. 
The loss function for the training of MDM and EMM in Retriever-Prompt is the in-batch contrastive loss. Specifically, given the queries $\textbf{Q}$ and their corresponding positive documents $\textbf{D}$ in a mini-batch during training, the negative documents for query $q_i \in \textbf{Q}$ are the positive documents of other queries in $\textbf{D}$. Given $\textbf{B}=\{q_i, d_i^+, d_{i,1}^-, d_{i,2}^-,..., d_{i,n}^-\}$, the input texts to PLM can be constructed as Equ.(\ref{template}):
\begin{equation}
    \begin{aligned}
    x_{q_{i}} &= [CLS]\,P_1^{t}\,q_i\,P_q^{t}\,[MASK_{q_{i}}], \\
    x_{d_{i}^+} &= [CLS]\,P_2^{t}\,d_{i}^+\,P_d^{t}\,[MASK_{d_{i}^+}],\\
    x_{d_{i,1}^-} &= [CLS]\,P_2^{t}\,d_{i,1}^-\,P_d^{t}\,[MASK_{d_{i,1}^-}],\\
    &... \\
   x_{d_{i,n}^-} &= [CLS]\,P_2^{t}\,d_{i,n}^-\,P_d^{t}\,[MASK_{d_{i,n}^-}], \\
   \end{aligned} \nonumber
\end{equation}
where $P_1^{t}$, $P_2^{t}$, $P_q^{t}$ and $P_d^{t}$ are obtained from Equ.(\ref{prompt encoder}). The representation of the text $x_i$ is $[MASK_i]$ obtained from the output hidden states of PLM. The similarity between $q$ and $d$ can be defined as:
\begin{equation}
    sim(q,d) = [MASK_q]^T[MASK_d].\nonumber
\end{equation}
The loss function for $\textbf{B}$ is:
\begin{equation}
        L(q_i, d_i^+, d_{i,1}^-, d_{i,2}^-,..., d_{i,n}^-) \\ 
        = -log{{e^{sim(q_i,d_i^+)}} \over {e^{sim(q_i,d_i^+)}+\sum_{j=1}^{n}{e^{sim(q_i,d_{i,j}^-)}}}},
        \label{loss} \nonumber
\end{equation}
and the total loss function $\mathcal{L}$ is: 
\begin{equation}
       \mathcal{L} = \sum_{i=1}^{m}L(q_i, d_i^+, d_{i,1}^-, d_{i,2}^-,..., d_{i,n}^-), \\ 
       \label{total_loss} \nonumber
\end{equation}
where m is the number of queries. The optimization of trainable parameters $\theta^*$ is:
\begin{equation}
\theta^* = \mathop {argmin}_{\theta}(\mathcal{L}).
\nonumber
\end{equation}
It is worth noting that in the training of matching description module, the trainable component is the prompt encoder and PLM is frozen, in the training of essential matching module, the trainable component is PLM and the prompt encoder is frozen. 

The loss function for the training of MDM and EMM in Reranker-Prompt is cross entropy loss. Specifically, let $\mathcal{M}$ be the pre-trained language model. The vocabulary of $\mathcal{M}$ is $\mathbb{V}$. $\mathbb{Y}$ is the label set in this task and $y$ is each label. We can design a verbalizer to map the label to the word in the vocabulary and mark it $v$. If the [MASK] is filled by “yes”, the label is $1$, if it is filled by “no”, the label is $0$. $\mathcal{M}(w|x_{input})$ is used to represent the probability of filling [MASK] by $w\in\mathbb{V}$. Relationship between $x_{input}$ and $y$ is modeled as:
\begin{equation}
s(y|x_{input}) = \mathcal{M}(v(y)|x_{input}), \nonumber
\end{equation}
\begin{equation}
P(y|x_{input}) = \frac{e^{s(y|x_{input})}}{\sum\nolimits_{y' \in \mathbb{Y}}e^{s(y'|x_{input})}}.
\nonumber
\end{equation}
The loss function of our method is cross entropy:
\begin{equation}
\mathcal{L} = -\sum_{i=1}^{N}[y_{i}\log P_{i}+(1-y_{i}) \log (1-P_{i})].
\nonumber
\end{equation}
Same as the training for retrieval, in the training of matching description module, the trainable component is the prompt encoder and PLM is frozen, in the training of essential matching module, the trainable component is PLM and the prompt encoder is frozen.

After training, we can get the matching function $F$, and the matching process can be simplified to:
\begin{equation}
y = F(x_1,x_2),
\nonumber
\end{equation}
$y \in \{0,1\}$, indicates whether $x_1$ and $x_2$ match. In ranking tasks such as DR, QA and RD, we use the probability that the word of [MASK] is predicted to be “yes” minus the probability that it is predicted to be “no” as the relevance between texts:
\begin{equation}
REL = \mathcal{M}(yes|x_{input})-\mathcal{M}(no|x_{input}).
\nonumber
\end{equation}
\section{Experiments}
In this section, we introduce the experiment settings and the performance of NIR-Prompt on the retrieval stage, reranking stage, and entire neural information retrieval pipeline.
\subsection{Experimental Settings}
\subsubsection{Research Questions}
\begin{figure}
  \centering
  \includegraphics[width=0.7\textwidth]{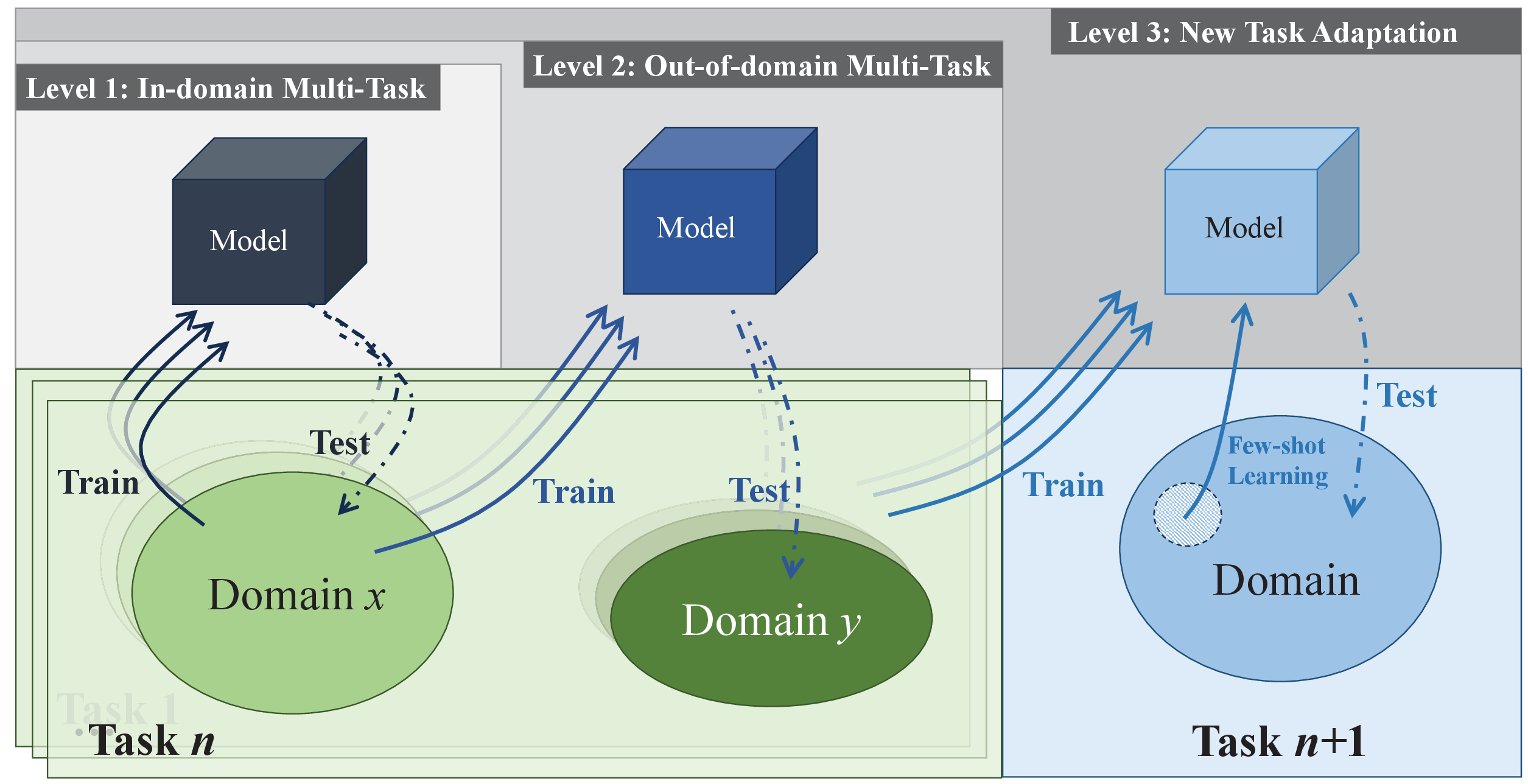}  \caption{Three levels of multi-task generalization.}
  \label{figure:3-level}
\end{figure}

To evaluate the generalization ability of NIR models at a fine-grained level, we define multi-task generalization into three levels. Figure~\ref{figure:3-level} describes the three levels in detail, which indicates the generalization across datasets, domains, and tasks. They can be evaluated by in-domain multi-task performance, out-of-domain multi-task performance, and new task adaptation respectively.  Based on these levels, we propose three corresponding research questions on the generalization of NIR models and the following experiments will answer these questions below.

\begin{itemize}
\item \textit{In-domain Multi-task Performance:} 
Can the information of different retrieval tasks be shared and facilitate each task? \textbf{Train:} Given mixed datasets of multiple tasks, the model performs multi-task learning on the training set. \textbf{Evaluate:} Test its performance on the testing set of each dataset. 
\item \textit{Out-of-domain Multi-task Performance:} Can our method capture generalizable essential matching signals and avoid fitting domain-specific biases? \textbf{Train:} Given mixed datasets of multiple tasks, the model performs multi-task learning on the training set. \textbf{Evaluate:} Testing the zero-shot learning ability of the multi-task model on out-of-domain datasets.
\item \textit{New Task Adaptation:} Can essential matching signals facilitate new task adaptation in low-resource scenarios? \textbf{Train:} We use leave-one-out to perform few-shot learning on new tasks. \textbf{Evaluate:} Test its performance on the testing set of each dataset. 
\end{itemize}
\begin{table}[t]
\setlength\tabcolsep{18pt}
\caption{Experiment results in each setting}
\begin{tabular}{llll}
\toprule
Setting             & Retrieval                                            & Reranking                                            & Pipeline                                                  \\ \hline
In-domain           & Table~\ref{table:basic-results retrieval} & Table~\ref{table:basic-results reranking} & Table~\ref{table:basic-results pipeline} \\
Out-of-domain       & Table~\ref{table:out-of-domain retrieval} & Table~\ref{table:out-of-domain reranking} & Table~\ref{table:out-of-domain pipeline} \\
New Task Adaptation & Table~\ref{few-shot retrieval} & Table~\ref{few-shot reranking} & Table~\ref{few-shot pipeline} \\ \toprule
\end{tabular}
\label{table:exp_in_setting}
\end{table}
\subsubsection{Datasets}
The datasets for the experiments consisting of three parts are shown in Table~\ref{table:datasets}. The first part is the six datasets selected from five tasks including document retrieval (DR), open-domain question answering (QA), retrieval-based dialogue (RD), paraphrase identification (PI), and natural language inference (NLI) (shown in Table~\ref{table:basic-datasets retrieval} and Table~\ref{table:basic-datasets reranking}). This part is used to train the NIR-Prompt model and evaluate the in-domain multi-task performance and new task adaptation. The second part is the eleven datasets selected from the same tasks as the first part, which is used to evaluate the out-of-domain multi-task performance. The third part is BEIR~\cite{beir}, a public benchmark to evaluate the zero-shot ability of retrieval models. As for the tasks in the retrieval and reranking stages, the former considers the information retrieval tasks including document retrieval (DR), open-domain question answering (QA), retrieval-based dialogue (RD), and BEIR, the latter considers these information retrieval tasks as well as the sentence pair classification tasks including paraphrase identification (PI) and natural language inference (NLI). {It is because although PI and NLI are not IR tasks, their modeling on the relationship between texts can also provide valuable matching signals. PI and NLI are only considered in reranking because they are the sentence-pair classification
tasks rather than retrieving the target texts from a large corpus, which is a gap with the in-batch training method in dual encoder architecture in the retrieval stage. In addition, mainstream studies~\cite{adapterfusion,hyperformer,attempt} perform PI and NLI based on the cross-encoder architecture in reranking, so we follow them and only use PI and NLI in reranking.}

\begin{table}[t]
\setlength\tabcolsep{15pt}
  \caption{Details of mixed training datasets for dense retrieval.}
  \label{table:basic-datasets retrieval}
  \scalebox{1}{
  \begin{tabular}{llll}
    \toprule
    Task & Dataset & Train (sampled) & Test\\
    \midrule
    DR & MQ2007 (MQ07)& 40,000 (queries) & 338 (queries)\\
    QA & TriviaQA (Trivia)& 40,000 (questions) & 8,837 (questions)\\
    RD & DailyDialog (DG)& 40,000 (dialogues)& 3,209 (dialogues)\\
  \bottomrule
\end{tabular}
}
\end{table}

\begin{table}[t]
\setlength\tabcolsep{15pt}
  \caption{Details of mixed training datasets for reranking.}
  \label{table:basic-datasets reranking}
  \scalebox{1}{
  \begin{tabular}{llll}
    \toprule
    Task & Dataset & Train (sampled) & Test\\
    \midrule
    DR & MQ2007 (MQ07)& 40,000 (q-d pairs) & 338 (queries)\\
    QA & TriviaQA (Trivia)& 40,000 (q-a pairs) & 8,837 (questions)\\
    RD & DailyDialog (DG)& 40,000 (dialogue pairs)& 3,209 (dialogues)\\
    PI & QQP+MSRP& 35,924+4,076 & 50,016+1,752\\
    NLI& MultiNLI (MNLI)& 40,000 & 39,296\\
  \bottomrule
\end{tabular}
}
\end{table}

\begin{table}[t]
\setlength\tabcolsep{1.5pt}
  \caption{Datasets in experiments. Level 1, 2 and 3 correspond to in-domain multi-task, out-of-domain multi-task and new task adaptation.}
  \label{table:datasets}
  \scalebox{0.9}{
  \begin{tabular}{cllcccc}
    \toprule
    Part & Task & Dataset & Retrieval Stage & Reranking Stage & Training & Evaluating \\
    \midrule
    \multirow{6}{*}{1} & DR  & MQ2007~\cite{mq2008} & \checkmark & \checkmark & \checkmark & Level 1, Level 3 \\
    & QA  & TriviaQA~\cite{trivia} & \checkmark & \checkmark & \checkmark & Level 1, Level 3 \\
    & RD  & DailyDialog~\cite{daily} & \checkmark & \checkmark & \checkmark & Level 1, Level 3 \\
    & PI  & QQP\tablefootnote{https://www.kaggle.com/c/quora-question-pairs}& & \checkmark & \checkmark & Level 1, Level 3 \\
    & PI  & MSRP~\cite{mrpc}&  & \checkmark & \checkmark & Level 1, Level 3 \\
    & NLI  & MultiNLI (MNLI)~\cite{mnli}&  & \checkmark & \checkmark & Level 1, Level 3 \\ \hdashline
    \multirow{12}{*}{2} & DR  & ClueWeb09-B (CW)~\cite{clueweb} & \checkmark & \checkmark &  & Level 2 \\
    & DR  & Robust04 (RB04)~\cite{robust04} & \checkmark & \checkmark &  & Level 2 \\
    & DR  & Gov2\tablefootnote{https://ir-datasets.com/gov2.html} & \checkmark & \checkmark &  & Level 2 \\
    & QA  & CuratedTREC (Trec)~\cite{trec} & \checkmark & \checkmark &  & Level 2 \\
    & QA  & Natural Questions (NQ)~\cite{nq} & \checkmark & \checkmark &  & Level 2 \\
    & QA  & WebQuestions (WQ)~\cite{webq} & \checkmark & \checkmark &  & Level 2 \\
    & RD  & Reddit & \checkmark & \checkmark &  & Level 2 \\
    & RD  & AmazonQA (AQ)~\cite{amazonqa} & \checkmark & \checkmark &  & Level 2 \\
    & NLI  & SNLI~\cite{snli} &  & \checkmark &  & Level 2 \\
    & NLI  & SICK-E~\cite{sick} &  & \checkmark &  & Level 2 \\
    & PI  & PARADE~\cite{parade} &  & \checkmark &  & Level 2 \\
    & PI  & TwitterURL (TURL)~\cite{twitter} &  & \checkmark &  & Level 2 \\ \hdashline
  3  & Multiple &  BEIR~\cite{beir} & \checkmark & \checkmark &  & Level 2 \\
  \bottomrule
\end{tabular}
}
\end{table}

\subsubsection{{Implementation Details}}
{We use BERT-base (109M) model~\footnote{https://huggingface.co/bert-base-uncased/tree/main} as the pre-trained model in experiments. In the training of EMM and MDM, for retrieval, we use the gradient descent method Adam with learning rate $10^{-5}$ and batch size $32$ to train the model, for reranking, the batch size is $15$. We train the models for $20$ epochs and evaluate the performance after each epoch. When there is no improvement in performance for 3 consecutive epochs, the training stops early. The lengths for continuous prompt are $6$, $6$, $5$, $5$ for $P_1^{t}$, $P_2^{t}$, $P_q^{t}$ and $P_d^{t}$ respectively. As for the data arrangement in the mixed datasets, for retrieval, the data in a batch comes from the same task because we use in-batch contrastive loss in training. The batches of each task are arranged alternately. For reranking, the amount of data from each task in a batch is sampled in balance. The training and inference are performed on a Tesla V100 32GB GPU.
}

\subsubsection{Special Pre-processing}
\begin{itemize}
\item Training and evaluation for retrieval. For QA, we construct q-a pairs in the training set as described in DPR~\cite{dpr}. For RD, we take the two adjacent texts in the dialogue as positive samples. For DR, we use the labeled sample pairs in the dataset as positive samples. The negative samples for these three tasks are obtained from the in-batch sampling, which are the positive samples for other queries. In evaluation, for each query of QA and DR, we use the set of labeled positive and negative samples in the dataset as its candidate list. For each query of RD, we sample $50$ samples as its candidate list.

In evaluation, the candidate document set for each query is the entire corpus.
\item Training and evaluation for reranking. For QA, RD, and DR, the construction method of positive samples is the same as that of retrieval. As for the negative samples, for QA and DR, the negative samples are obtained from the labeled text pairs from the datasets, for RD, the negative samples are randomly sampled from the corpus. For PI, the size of public datasets is small, so we select two datasets including MSRP and QQP. For NLI, we convert this task into a binary classification task that whether hypotheses can be inferred from the premises. In evaluation, we use BM25 to retrieve items for each query and use the reranking model to rerank them. Specifically for DR, the candidate document set for each query is the labeled documents obtained from the dataset, which is much smaller than the entire corpus.
\item Evaluation for information retrieval pipeline. For the evaluation of the neural information retrieval pipeline, we use NIR-Prompt and traditional fine-tuning paradigm to retrieve the candidate document set based on the entire corpus and rerank them respectively.
\end{itemize}

\subsubsection{Baselines and Measures}
Our baselines consist of task-specific and multi-task models for dense retrieval and reranking trained by the traditional fine-tuning paradigm.

The task-specific models (i.e. \textbf{Fine-tuning$_{sp}$}) are specifically trained on the dataset corresponding to each task listed in Table~\ref{table:basic-datasets retrieval} and Table~\ref{table:basic-datasets reranking}, and tested on the corresponding task.  For dense retrieval, the implementation details of Fine-tuning$_{sp}$ are consistent with DPR. For reranking, the implementation details of Fine-tuning$_{sp}$ are consistent with MonoBert. {DPR uses [CLS] token of BERT to represent query and document as dense vectors and use the in-batch contrastive loss as the learning objective. MonoBer concatenates query and document as the input and feeds the embedding of [CLS] into a feed-forward network to judge the relevance. Both of them are based on the traditional fine-tuning paradigm of PLMs. In this paper, we only discuss the advantages of NIR-Prompt over the fine-tuning paradigm. More optimization techniques such as hard negative sampling and distillation are not specifically discussed, because they can also be applied to our method.}

In order to capture the essential matching signals, our method needs to be mixed-trained on multiple datasets. Therefore, for a fair comparison, we also use a multi-task training method for the traditional fine-tuning method on multiple datasets. Specifically, for the fine-tuning method in dense retrieval and reranking (DPR and MonoBert), we introduce two multi-task training methods. \textbf{Fine-tuning$_{multi}$}, which uses the traditional fine-tuning paradigm to train PLM on the mixed datasets without any task-specific marks. \textbf{Fine-tuning$_{mark}$}, which adds the task-specific marks to the input text to be used as task differentiation in multi-task training. Besides, the following are the multi-task training methods specifically for reranking worth to be considered. \textbf{MT-DNN}~\cite{mt-dnn} adds the task-specific feed-forward networks for each task, and we reproduce it on our mixed datasets. This method introduces additional parameters, and the number of parameters increases with the number of tasks, but NIR-Prompt does not need any task-specific layers during inference. There is also a multi-task learning framework that converts each task into a unified question-answering format~\cite{zero-shot-tc,tc,MQAN,t5}. {We reproduce this framework on our mixed datasets using T5-base (220M)\footnote{https://huggingface.co/google/t5-v1\_1-base} and call them \textbf{MTL$_{T5}$}}. Since T5 has been pre-trained on multiple supervised downstream tasks that will be tested in our experiment, which is unfair for comparison, we choose T5 1.1, which is only pre-trained on unsupervised datasets. \textbf{BM25}~\cite{bm25} also has strong multi-task generalization ability, we use it as one of the baselines. {We also use the parameter-efficient multi-task training and few-shot learning methods such as \textbf{HyperFormer}~\cite{hyperformer}, \textbf{ATTEMPT}~\cite{attempt} and \textbf{AdapterFusion}~\cite{adapterfusion} as the baselines. We reproduce them on our datasets.} In multi-task training, there are some tricks about data sampling, loss construction, and task scheduling, which can be used by both baselines and NIR-Prompt, so we do not compare them in detail. 

In this paper, we want to show that the NIR model trained with NIR-Prompt (based on the idea of decoupling the process of signal capturing and signal combination) can get better multi-task generalization ability than the traditional fine-tuning paradigm. Some methods~\cite{monobert,plminir,qa_bert_sota,pi-bert,match-ignition,cedr,ance,condenser} are not given special consideration because they are just variants based on this fine-tuning paradigm, which are inconsistent with our motivation and can also be incorporated into our method. {We train \textbf{RetroMAE}~\cite{retromae} (a state-of-the-art IR model on both in-domain and out-of-domain generalization) by our method and the traditional fine-tuning paradigm respectively on the mixed datasets to show the compatibility of our method to SOTA IR Models.}

For our method, \textbf{Retriever-Prompt$_{mp}$}, \textbf{Retrieve-Prompt$_{cp}$} correspond to manual and continuous prompt for dense retrieval. \textbf{Reranker-Prompt$_{mp}$}, \textbf{Reranker-Prompt$_{cp}$} and \textbf{Reranker-Prompt$_{hp}$} correspond to manual, continuous and hybrid prompt for reranking.

As for the metrics for evaluation, we use Accuracy and F1-score to evaluate NLI and PI~\cite{esim}. Accuracy aims to measure the proportion of samples that the reranking model correctly classifies the matching relationship between two texts. F1-score is the combination of recall (the distinguishing ability of positive samples) and precision (the distinguishing ability of positive samples) and reflects the robustness of the model. For QA and RD, these two tasks usually only return the first-ranked sample to the user, so P@1 and MRR are suitable for them~\cite{SRNN}. P@1 measures the proportion of queries for which the correct answer is ranked first. MRR measures the position of the first correct answer in the ranked list returned by the model. For DR, we use NDCG~\cite{prop} that can measure the relevance of documents and ranking position in the returned list.

\subsection{Main Results}

\subsubsection{In-domain Multi-task Performance} \label{in-domain}
We do multi-task mixed training for retrieval and reranking on Table~\ref{table:basic-datasets retrieval} and Table~\ref{table:basic-datasets reranking} respectively and test their performance on the corresponding test dataset. The results are shown in Table~\ref{table:basic-results retrieval}, ~\ref{table:basic-results reranking} and~\ref{table:basic-results pipeline}. For retrieval, reranking, and the entire neural information pipeline, NIR-Prompt shows better performance than the traditional fine-tuning paradigm. Compare multi-task models with the task-specific model (Fine-tuning$^{1}_{sp}$), the performance of fine-tuning paradigm drops significantly but our method surpasses it, which shows that in the in-domain setting, the traditional end-to-end fine-tuning paradigm cannot effectively utilize the information shared between tasks to promote each other, but interfere with each other, and our method solves this problem. The experiment result can answer the first research question that NIR-Prompt can capture the shared information between tasks and exploit it to facilitate each other. This is mainly due to our idea of decoupling the process of signal capturing and signal combination, which can help the multi-task model distinguish and adapt to different tasks.

\begin{table*}[htbp]
  \caption{In-domain multi-task performance of the model trained on the mixed datasets listed in Table~\ref{table:basic-datasets retrieval} for dense retrieval and reranking. \textbf{Boldface} indicates the best results of multi-task models and the results over Fine-tuning$_{sp}$ are denoted as `$*$'. Results with significant performance improvement with p-value $ \leq 0.05$ compared with all multi-task models of baselines are denoted as `$+$'.}
  \subtable[In-domain retrieval]{
\renewcommand\arraystretch{1.1}

\setlength\tabcolsep{11pt}
\renewcommand{\cellset}{\renewcommand{\arraystretch}{0.5}} 

\centering
   \scalebox{0.9}{
\begin{tabular}{llllllllll}
\toprule
\multirow{2}{*}{Method} & \multicolumn{2}{c}{Trivia$_{QA}$} & \multicolumn{2}{c}{DailyDialog$_{RD}$} & MQ2007$_{DR}$ \\ 
                        &  P@1(\%)           & MRR(\%)             & P@1(\%)           & MRR(\%)         & NDCG@10(\%) \\ \hline  \hline 
\multicolumn{10}{c}{Task-specific model for each specific task} \\    
Fine-tuning$^{1}_{sp}$  & 46.92          & 54.70          & 64.19            & 70.42            & 21.36  \\  \hline  
\multicolumn{10}{c}{Multi-task model for all tasks} \\
BM25   & 46.88          & 54.67          & 35.81             & 50.39          & 15.21        \\

Fine-tuning$^{1}_{multi}$      & \makecell[l]{44.16}          & \makecell[l]{52.03}          & \makecell[l]{62.89}             & \makecell[l]{68.75}        & \makecell[l]{19.18} \\

Fine-tuning$^{1}_{mark}$      & \makecell[l]{44.11}          & \makecell[l]{51.98}          & \makecell[l]{63.47}             & \makecell[l]{68.80}              & \makecell[l]{20.46} \\
{ATTEMPT$^{1}$}       & \makecell[l]{{44.05}}          & \makecell[l]{{51.57}}          & \makecell[l]{{63.05}}             & \makecell[l]{{68.35}}          & \makecell[l]{{20.25}} \\
{HyperFormer$^{1}$}       & \makecell[l]{{44.23}}          & \makecell[l]{{52.09}}          & \makecell[l]{{63.98}}             & \makecell[l]{{69.10}}          & \makecell[l]{{20.75}} \\
{AdapterFusion$^{1}$}    & \makecell[l]{{44.15}}          & \makecell[l]{{52.07}}          & \makecell[l]{{63.91}}             & \makecell[l]{{68.95}}          & \makecell[l]{{20.30}} \\
\hdashline

Retriever-Prompt$_{mp}$     & \makecell[l]{48.31$^{*}_{+}$}          & \makecell[l]{57.62$^{*}_{+}$}          & \makecell[l]{65.37$^{*}_{+}$}             & \makecell[l]{71.02$^{*}_{+}$}             & 23.92$^{*}_{+}$  \\

Retriever-Prompt$_{cp}$     & \makecell[l]{\textbf{50.23$^{*}_{+}$}}          & \makecell[l]{\textbf{59.24$^{*}_{+}$}}          & \makecell[l]{\textbf{66.21$^{*}_{+}$}}             & \makecell[l]{\textbf{72.68$^{*}_{+}$}}             & \textbf{24.80$^{*}_{+}$}      \\
\toprule
\end{tabular}
}
  \label{table:basic-results retrieval}
}
\subtable[In-domain reranking]{
\renewcommand\arraystretch{1.1}

\setlength\tabcolsep{2pt}
\renewcommand{\cellset}{\renewcommand{\arraystretch}{0.5}} 

\centering
   \scalebox{0.85}{
\begin{tabular}{llllllllll}
\toprule
\multirow{2}{*}{Method} & \multicolumn{2}{c}{Trivia$_{QA}$} & \multicolumn{2}{c}{DailyDialog$_{RD}$} & \multicolumn{2}{c}{MSRP+QQP$_{PI}$}  & \multicolumn{2}{c}{MultiNLI$_{NLI}$} & MQ2007$_{DR}$ \\ 
                        &  P@1(\%)           & MRR(\%)             & P@1(\%)           & MRR(\%)               & Acc.(\%)           & F$_1$(\%)             & Acc.(\%)           & F$_1$(\%)            & NDCG@10(\%) \\ \hline  \hline 
\multicolumn{10}{c}{Task-specific model for each specific task} \\    
Fine-tuning$^{2}_{sp}$  & 47.17          & 55.62          & 78.59            & 84.86            & 83.42          & 79.66  & 84.40            & 77.35          & 50.60   \\  \hline  
\multicolumn{10}{c}{Multi-task model for all tasks} \\
MTL$_{T5}$      & \makecell[l]{36.24}          & \makecell[l]{46.63}          & \makecell[l]{62.49}             & \makecell[l]{74.99}            & \makecell[l]{71.54}          & \makecell[l]{74.70}         & \makecell[l]{77.28}            & \makecell[l]{64.73}           & \makecell[l]{48.63} \\

MT-DNN      & \makecell[l]{46.01}          & \makecell[l]{54.46}          & \makecell[l]{75.10}             & \makecell[l]{82.05}            & \makecell[l]{81.50}          & \makecell[l]{77.10}         & \makecell[l]{82.33}            & \makecell[l]{73.92}           & \makecell[l]{47.94} \\ 

Fine-tuning$^{2}_{multi}$      & \makecell[l]{46.19}          & \makecell[l]{54.95}          & \makecell[l]{74.29}             & \makecell[l]{81.47}            & \makecell[l]{80.85}          & \makecell[l]{77.13}         & \makecell[l]{81.45}            & \makecell[l]{72.77}           & \makecell[l]{47.59} \\

Fine-tuning$^{2}_{mark}$      & \makecell[l]{46.05}          & \makecell[l]{54.72}          & \makecell[l]{74.80}             & \makecell[l]{81.95}            & \makecell[l]{81.47}          & \makecell[l]{75.44}         & \makecell[l]{82.55}            & \makecell[l]{74.01}           & \makecell[l]{49.94} \\
{ATTEMPT$^{2}$}      & \makecell[l]{{45.21}}          & \makecell[l]{{53.85}}          & \makecell[l]{{73.76}}             & \makecell[l]{{80.83}}            & \makecell[l]{{81.59}}          & \makecell[l]{{76.74}}         & \makecell[l]{{82.87}}            & \makecell[l]{{74.35}}           & \makecell[l]{{49.63}} \\
{HyperFormer$^{2}$}      & \makecell[l]{{46.33}}          & \makecell[l]{{54.95}}          & \makecell[l]{{74.35}}             & \makecell[l]{{81.20}}            & \makecell[l]{{82.43}}          & \makecell[l]{{77.38}}         & \makecell[l]{{83.67}}            & \makecell[l]{{75.02}}           & \makecell[l]{{50.01}} \\
{AdapterFusion$^{2}$}      & \makecell[l]{{45.69}}          & \makecell[l]{{54.05}}          & \makecell[l]{{73.95}}             & \makecell[l]{{81.27}}            & \makecell[l]{{82.76}}          & \makecell[l]{{77.53}}         & \makecell[l]{{82.96}}            & \makecell[l]{{74.55}}           & \makecell[l]{{49.70}} \\
\hdashline

Reranker-Prompt$_{mp}$     & \makecell[l]{47.67$^{*}_{+}$}          & \makecell[l]{55.87$^{*}_{+}$}          & \makecell[l]{78.00$_{+}$}             & \makecell[l]{84.47$_{+}$}            & \makecell[l]{82.78$_{+}$} & \makecell[l]{79.48$_{+}$}          & \makecell[l]{83.56$_{+}$}           & \makecell[l]{76.07$_{+}$}         & 51.42$^{*}_{+}$  \\

Reranker-Prompt$_{cp}$     & \makecell[l]{47.79$^{*}_{+}$}          & \makecell[l]{55.98$^{*}_{+}$}          & \makecell[l]{78.01$_{+}$}             & \makecell[l]{84.58$_{+}$}            & \makecell[l]{83.36$_{+}$}  & \makecell[l]{79.42$_{+}$}          & \makecell[l]{84.56$^{*}_{+}$ }           & \makecell[l]{76.82$_{+}$}    & 51.52$^{*}_{+}$      \\

Reranker-Prompt$_{hp}$    & \makecell[l]{\textbf{47.92$^{*}_{+}$}}  & \makecell[l]{\textbf{56.14$^{*}_{+}$}}  & \makecell[l]{\textbf{78.65$^{*}_{+}$}}    & \makecell[l]{\textbf{84.87$^{*}_{+}$}}    & \makecell[l]{\textbf{83.55$^{*}_{+}$}}           & \makecell[l]{\textbf{79.82$^{*}_{+}$}}          & \makecell[l]{\textbf{84.64$^{*}_{+}$}}    & \makecell[l]{\textbf{77.40$^{*}_{+}$}}  & \makecell[l]{\textbf{52.01$^{*}_{+}$}} \\
\toprule
\end{tabular}}
\label{table:basic-results reranking}
}
\subtable[In-domain neural IR pipeline. FT is the traditional fine-tuning paradigm, RP$^{1}_{cp}$ is Retriever-Prompt$_{cp}$ and RP$^{2}_{hp}$ is Reranker-Prompt$_{cp}$.]{
\renewcommand\arraystretch{1.25}

\setlength\tabcolsep{12pt}
\renewcommand{\cellset}{\renewcommand{\arraystretch}{0.5}} 

\centering
   \scalebox{0.9}{
\begin{tabular}{llllllllll}
\toprule
\multirow{2}{*}{Method} & \multicolumn{2}{c}{Trivia$_{QA}$} & \multicolumn{2}{c}{DailyDialog$_{RD}$} & MQ2007$_{DR}$ \\ 
                        &  P@1(\%)           & MRR(\%)             & P@1(\%)           & MRR(\%)         & NDCG@10(\%) \\ \hline  \hline 
\multicolumn{10}{c}{Task-specific model for each specific task} \\    
FT$^{1}_{sp}$+FT$^{2}_{sp}$  & 47.25          & 55.73          & 80.13            & 85.09            & 28.56  \\  \hline 
\multicolumn{10}{c}{Multi-task model for all tasks} \\

FT$^{1}_{mark}$+FT$^{2}_{mark}$      & \makecell[l]{45.81}          & \makecell[l]{54.04}          & \makecell[l]{76.33}             & \makecell[l]{82.84}              & \makecell[l]{27.46} \\
{HF$^{1}$+HF$^{2}$}     & \makecell[l]{{46.02}}          & \makecell[l]{{54.79}}          & \makecell[l]{{76.85}}             & \makecell[l]{{83.44}}          & \makecell[l]{{28.00}} \\ \hdashline
RP$^{1}_{cp}$+RP$^{2}_{hp}$     & \makecell[l]{\textbf{49.78$^{*}_{+}$}}          & \makecell[l]{\textbf{58.95$^{*}_{+}$}}          & \makecell[l]{\textbf{80.42$^{*}_{+}$}}             & \makecell[l]{\textbf{85.93$^{*}_{+}$}}             & \textbf{30.22$^{*}_{+}$}      \\
\toprule
\end{tabular}
}
  \label{table:basic-results pipeline}
}
\end{table*}

\subsubsection{Out-of-domain Multi-task Performance}
\label{out-of-domain}
We test the retrieval and reranking model trained on Table~\ref{table:basic-datasets retrieval} and Table~\ref{table:basic-datasets reranking} on their unseen datasets (Part 2 of Table~\ref{table:datasets}). The results shown in Table~\ref{table:out-of-domain retrieval},~\ref{table:out-of-domain reranking} and~\ref{table:out-of-domain pipeline} indicate that NIR-Prompt has better out-of-domain multi-task performance than traditional fine-tuning paradigm. Highly diverse tasks mixed training makes the model avoid overfitting the dataset bias and capture the essential matching signals that can be used across tasks and domains. The prompt tokens are used as the task description to guide the learning of essential matching signals during mixed training and adapt these signals to different tasks and domains, which is a more reasonable approach to conforming the knowledge in PLM. These factors improve the generalization performance of NIR-Prompt. As for the traditional fine-tuning paradigm, the performance of the multi-task model drops more seriously than the task-specific model Fine-tuning$_{sp}$, which indicates that the multi-task model trained by fine-tuning paradigm cannot distinguish each task well and tasks interfere with each other. On the contrary, NIR-Prompt has the stronger task-distinguishing ability and enables tasks to utilize shared information to promote each other. The experiment result can answer the second research question that NIR-prompt can capture the essential matching signals that can be used across domains to improve the out-of-domain generalization ability.

\begin{table*}[htbp]
  \caption{Out-of-domain performance of the model on unseen datasets listed in Part 2 of Table~\ref{table:datasets} for dense retrieval and reranking. \textbf{Boldface} indicates the best results of multi-task models and the results over Fine-tuning$_{sp}$ are denoted as `$*$'. Results with significant performance improvement with p-value $ \leq 0.05$ compared with all multi-task models of baselines are denoted as `$+$'. }
\subtable[Out-of-domain retrieval]{
\renewcommand\arraystretch{1.45}

\setlength\tabcolsep{1.8pt}
\renewcommand{\cellset}{\renewcommand{\arraystretch}{1.5}} 

\centering
   \scalebox{0.75}{
\begin{tabular}{llllllllllllllllllllllll}
\toprule
\multirow{2}{*}{Method}  & \multicolumn{2}{c}{Reddit$_{RD}$} & \multicolumn{2}{c}{AQ$_{RD}$} & \multicolumn{2}{c}{Trec$_{QA}$} & \multicolumn{2}{c}{NQ$_{QA}$} & \multicolumn{2}{c}{WQ$_{QA}$} & CW$_{DR}$      & RB04$_{DR}$  & Gov2$_{DR}$ \\ 
                         & P@1           & MRR  & P@1           & MRR    & P@1           & MRR             & P@1           & MRR  & P@1           & MRR  & \multicolumn{3}{c}{NDCG@10}\\ \hline \hline 
\multicolumn{21}{c}{Task-specific model for each specific task} \\
Fine-tuning$^{1}_{sp}$            &32.45 &50.31 & 74.14 & 83.42 & 36.50          & 50.09          & 20.85         & 32.19          & 26.52             & 39.05           & 22.75         & 27.45 & 33.32 \\  \hline
\multicolumn{21}{c}{Multi-task model for all tasks } \\   
BM25             & 24.63 & 44.81 & 69.73 & 76.08 & 31.94          & 46.18      & 24.75$^{*}$    & 37.30$^{*}$ &  26.47$^{*}$          & 39.17$^{*}$            & 22.58         & 30.12$^{*}$ & 38.01$^{*}$ \\ 

Fine-tuning$^{1}_{multi}$            & \makecell[l]{36.15$^{*}$}  &\makecell[l]{52.78$^{*}$}   
&\makecell[l]{71.01}  &\makecell[l]{78.61}  & \makecell[l]{32.55}          & \makecell[l]{48.21}           & \makecell[l]{20.01}          & \makecell[l]{31.79}          & \makecell[l]{25.02}             & \makecell[l]{36.89}           & \makecell[l]{22.85$^{*}$}                   & \makecell[l]{26.12} & \makecell[l]{35.97$^{*}$} \\

Fine-tuning$^{1}_{mark}$              &\makecell[l]{37.48$^{*}$}  &\makecell[l]{53.61$^{*}$}   &\makecell[l]{72.36}  &\makecell[l]{81.27} & \makecell[l]{32.81}          & \makecell[l]{48.76}  & \makecell[l]{20.31}          & \makecell[l]{32.05}          & \makecell[l]{25.52}             & \makecell[l]{37.25}           & \makecell[l]{22.97$^{*}$}                   & \makecell[l]{26.26} & \makecell[l]{36.11$^{*}$}  \\
{ATTEMPT$^{1}$}  &\makecell[l]{{37.52$^{*}$}}   &\makecell[l]{{53.79$^{*}$}}  &\makecell[l]{{72.45}} & \makecell[l]{{81.33}}          & \makecell[l]{{32.96}}  & \makecell[l]{{49.03}}          & \makecell[l]{{20.45}}          & \makecell[l]{{32.35}}             & \makecell[l]{{25.68}}           & \makecell[l]{{37.40}}                  & \makecell[l]{{23.05$^{*}$}} & \makecell[l]{{26.53}}  & \makecell[l]{{36.29$^{*}$}}\\

{HyperFormer$^{1}$}   &\makecell[l]{{38.05$^{*}$}}   &\makecell[l]{{54.09$^{*}$}}  &\makecell[l]{{73.01}} & \makecell[l]{{81.96}}          & \makecell[l]{{33.15}}  & \makecell[l]{{49.25}}          & \makecell[l]{{20.98}}          & \makecell[l]{{32.75}}             & \makecell[l]{{26.03}}           & \makecell[l]{{37.82}}                  & \makecell[l]{{23.18$^{*}$}} & \makecell[l]{{26.87}}  & \makecell[l]{{36.92$^{*}$}}\\

{AdapterFusion$^{1}$}    &\makecell[l]{{37.50$^{*}$}}   &\makecell[l]{{53.75$^{*}$}}  &\makecell[l]{{72.96}} & \makecell[l]{{81.52}}          & \makecell[l]{{32.78}}  & \makecell[l]{{48.82}}          & \makecell[l]{{20.65}}          & \makecell[l]{{32.47}}             & \makecell[l]{{25.50}}           & \makecell[l]{{37.32}}                  & \makecell[l]{{22.96$^{*}$}} & \makecell[l]{{26.49}}  & \makecell[l]{{36.17$^{*}$}}\\
 \hdashline

Retriever-Prompt$^{1}_{mp}$    & 38.03$^{*}_{+}$ & 53.86$^{*}_{+}$ & 77.54$^{*}_{+}$ & 85.64$^{*}_{+}$  & \makecell[l]{36.47$_{+}$}          & \makecell[l]{50.02$_{+}$}          & \makecell[l]{25.01$^{*}_{+}$}          & \makecell[l]{38.65$^{*}_{+}$}          & \makecell[l]{24.71}             & \makecell[l]{36.42}           & \makecell[l]{26.32$^{*}_{+}$}  & \makecell[l]{30.10$^{*}_{+}$} & \makecell[l]{38.23$^{*}_{+}$}       \\

Retriever-Prompt$^{1}_{cp}$    &\textbf{40.52}$^{*}_{+}$ &\textbf{56.16}$^{*}_{+}$   & \textbf{82.16}$^{*}_{+}$ & \textbf{87.32}$^{*}_{+}$ & \makecell[l]{\textbf{38.46}$^{*}_{+}$}          & \makecell[l]{\textbf{52.52}$^{*}_{+}$ }          & \makecell[l]{\textbf{25.58}$^{*}_{+}$ }           & \makecell[l]{\textbf{39.79}$^{*}_{+}$}          & \makecell[l]{\textbf{26.82}$^{*}_{+}$}              & \makecell[l]{\textbf{39.33}$^{*}_{+}$}              & \makecell[l]{\textbf{27.37}$^{*}_{+}$}  & \makecell[l]{\textbf{32.21}$^{*}_{+}$} & \makecell[l]{\textbf{39.28}$^{*}_{+}$} \\

\toprule
\end{tabular}
  \label{table:out-of-domain retrieval}
}
}
\subtable[Out-of-domain reranking]{
\renewcommand\arraystretch{1.5}

\setlength\tabcolsep{1.25pt}
\renewcommand{\cellset}{\renewcommand{\arraystretch}{0.5}} 

\centering
   \scalebox{0.52}{
\begin{tabular}{llllllllllllllllllllllll}
\toprule
\multirow{2}{*}{Method} & \multicolumn{2}{c}{SNLI$_{NLI}$} &\multicolumn{2}{c}{SICK-E$_{NLI}$} & \multicolumn{2}{c}{PARADE$_{PI}$}   & \multicolumn{2}{c}{TURL$_{PI}$} & \multicolumn{2}{c}{Reddit$_{RD}$} & \multicolumn{2}{c}{AQ$_{RD}$} & \multicolumn{2}{c}{Trec$_{QA}$} & \multicolumn{2}{c}{NQ$_{QA}$} & \multicolumn{2}{c}{WQ$_{QA}$} & CW$_{DR}$      & RB04$_{DR}$  & Gov2$_{DR}$ \\ 
                        & Acc.           & F$_1$             & Acc.           & F$_1$           & Acc.           & F$_1$  & Acc.           & F$_1$ & P@1           & MRR  & P@1           & MRR    & P@1           & MRR             & P@1           & MRR  & P@1           & MRR  & \multicolumn{3}{c}{NDCG@10}\\ \hline \hline 
\multicolumn{21}{c}{Task-specific model for each specific task} \\
Fine-tuning$^{2}_{sp}$            & 82.19          & 74.21          & 81.61            & 73.12           & 68.34         & 60.85   &83.33 &60.45 &49.64
 &64.13 & 77.64 & 85.22 & 47.95          & 62.19          & 22.97          & 36.24          & 36.52             & 51.50           & 26.01         & 40.28 & 43.27 \\  \hline
\multicolumn{21}{c}{Multi-task model for all tasks } \\   

MTL$_{T5}$             & \makecell[l]{69.56}          & \makecell[l]{34.72}          & \makecell[l]{67.60}             &\makecell[l]{52.51}          & \makecell[l]{69.81$^{*}$}         & \makecell[l]{59.85}    &\makecell[l]{83.83$^{*}$} &\makecell[l]{62.53$^{*}$}  &\makecell[l]{33.96}  &\makecell[l]{50.35}   
&\makecell[l]{72.49}  &\makecell[l]{79.95}  & \makecell[l]{34.88}          & \makecell[l]{50.34}           & \makecell[l]{14.42}          & \makecell[l]{26.61}          & \makecell[l]{20.70}             & \makecell[l]{36.09}           & \makecell[l]{25.75}                   & \makecell[l]{39.90} & \makecell[l]{46.89$^{*}$} \\

MT-DNN         & \makecell[l]{77.85}          & \makecell[l]{70.30}          & \makecell[l]{79.11}             & \makecell[l]{70.54}           & \makecell[l]{65.10 }         & \makecell[l]{49.68 }  & 82.55 & 52.17 & 55.72$^{*}$ & 68.36$^{*}$ & 75.98 & 83.47  & \makecell[l]{41.55 }          & \makecell[l]{57.09 }          & \makecell[l]{21.27 }          & \makecell[l]{34.56}          & \makecell[l]{32.34}             & \makecell[l]{48.52}   & \makecell[l]{25.16} & \makecell[l]{40.18} & \makecell[l]{47.09$^{*}$}       \\

Fine-tuning$^{2}_{multi}$             & \makecell[l]{78.24}          & \makecell[l]{66.73}          & \makecell[l]{77.57}             &\makecell[l]{68.79}          & \makecell[l]{64.95}         & \makecell[l]{60.27}    &\makecell[l]{76.25} &\makecell[l]{50.47} &\makecell[l]{50.95$^{*}$}  &\makecell[l]{66.63$^{*}$}   
&\makecell[l]{73.09}  &\makecell[l]{80.21}  & \makecell[l]{45.55}          & \makecell[l]{60.12}           & \makecell[l]{22.22}          & \makecell[l]{36.01}          & \makecell[l]{32.27}             & \makecell[l]{48.49}           & \makecell[l]{26.05$^{*}$}                   & \makecell[l]{40.89$^{*}$} & \makecell[l]{47.05$^{*}$} \\

Fine-tuning$^{2}_{mark}$             & \makecell[l]{80.69}          & \makecell[l]{{71.45}}          & \makecell[l]{79.06 }             &\makecell[l]{67.81}          & \makecell[l]{67.67}         & \makecell[l]{60.71}    &\makecell[l]{81.39} &\makecell[l]{59.06} &\makecell[l]{56.20$^{*}$}  &\makecell[l]{69.75$^{*}$}   
&\makecell[l]{76.85}  &\makecell[l]{84.66} & \makecell[l]{45.82}          & \makecell[l]{60.71}  & \makecell[l]{25.81$^{*}$}          & \makecell[l]{39.82$^{*}$}          & \makecell[l]{34.82}             & \makecell[l]{50.17}           & \makecell[l]{25.56}                   & \makecell[l]{40.21} & \makecell[l]{47.17$^{*}$}  \\
{ATTEMPT$^{2}$}                 & \makecell[l]{{81.05}}          & \makecell[l]{{{71.78}}}          & \makecell[l]{{79.53}}             &\makecell[l]{{68.04}}          & \makecell[l]{{67.75}}         & \makecell[l]{{60.96}}    &\makecell[l]{{81.68}} &\makecell[l]{{59.45}} &\makecell[l]{{56.77}}  &\makecell[l]{{70.03}}   
&\makecell[l]{{77.32}}  &\makecell[l]{{85.34}} & \makecell[l]{{46.02}}          & \makecell[l]{{61.35}}  & \makecell[l]{{25.85}}          & \makecell[l]{{39.87}}          & \makecell[l]{{34.98}}             & \makecell[l]{{50.56}}           & \makecell[l]{{26.13}}                   & \makecell[l]{{40.76}} & \makecell[l]{{47.23}} \\
{HyperFormer$^{2}$}     & \makecell[l]{{81.93}}          & \makecell[l]{{{72.31}}}          & \makecell[l]{{80.15}}             &\makecell[l]{{68.60}}          & \makecell[l]{{68.04}}         & \makecell[l]{{61.23}}    &\makecell[l]{{82.05}} &\makecell[l]{{60.03}} &\makecell[l]{{57.32}}  &\makecell[l]{{70.40}}   
&\makecell[l]{{78.08}}  &\makecell[l]{{85.91}} & \makecell[l]{{46.75}}          & \makecell[l]{{61.90}}  & \makecell[l]{{25.91}}          & \makecell[l]{{39.90}}          & \makecell[l]{{35.10}}             & \makecell[l]{{50.73}}           & \makecell[l]{{26.21}}                   & \makecell[l]{{40.85}} & \makecell[l]{{47.48}} \\
{AdapterFusion$^{2}$}      & \makecell[l]{{81.12}}          & \makecell[l]{{{71.83}}}          & \makecell[l]{{79.66}}             &\makecell[l]{{68.20}}          & \makecell[l]{{68.75}}         & \makecell[l]{{61.22}}    &\makecell[l]{{82.70}} &\makecell[l]{{60.38}} &\makecell[l]{{56.53}}  &\makecell[l]{{69.91}}   
&\makecell[l]{{77.12}}  &\makecell[l]{{85.03}} & \makecell[l]{{46.82}}          & \makecell[l]{{62.01}}  & \makecell[l]{{26.03}}          & \makecell[l]{{39.98}}          & \makecell[l]{{35.17}}             & \makecell[l]{{50.72}}           & \makecell[l]{{26.08}}                   & \makecell[l]{{40.56}} & \makecell[l]{{47.19}} \\
 \hdashline

Reranker-Prompt$_{mp}$     & \makecell[l]{82.01$_{+}$}          & \makecell[l]{72.71$_{+}$ }          & \makecell[l]{79.14}             & \makecell[l]{70.50}           & \makecell[l]{69.24$^{*}$}         & \makecell[l]{62.15$^{*}$}  & 84.82$^{*}_{+}$ & 63.43$^{*}_{+}$ & 57.30$^{*}_{+}$ & 70.95$^{*}_{+}$ & 79.40$^{*}_{+}$ & 86.24$^{*}_{+}$  & \makecell[l]{48.89$^{*}_{+}$}          & \makecell[l]{63.02$^{*}_{+}$}          & \makecell[l]{\textbf{26.21$^{*}_{+}$} }          & \makecell[l]{\textbf{40.09$^{*}_{+}$}}          & \makecell[l]{35.38$_{+}$}             & \makecell[l]{50.32$_{+}$}           & \makecell[l]{\textbf{26.93$^{*}_{+}$}}  & \makecell[l]{41.29$^{*}_{+}$} & \makecell[l]{\textbf{48.53$^{*}_{+}$}}       \\

Reranker-Prompt$_{cp}$     & \makecell[l]{82.06$_{+}$}          & \makecell[l]{73.91$_{+}$}          & \makecell[l]{81.53$_{+}$}              & \makecell[l]{73.04$_{+}$}             & \makecell[l]{70.01$^{*}_{+}$}         & \makecell[l]{62.20$^{*}_{+}$}    & 84.75$^{*}_{+}$ & 63.98$^{*}_{+}$ &57.70$^{*}_{+}$ &71.96$^{*}_{+}$   & 83.26$^{*}_{+}$ & 88.35$^{*}_{+}$ & \makecell[l]{48.06$^{*}_{+}$}          & \makecell[l]{62.37$^{*}_{+}$ }          & \makecell[l]{23.50$^{*}$ }           & \makecell[l]{36.81$^{*}$}          & \makecell[l]{35.90$_{+}$}              & \makecell[l]{51.31$_{+}$}              & \makecell[l]{26.41$^{*}_{+}$}  & \makecell[l]{40.54$^{*}$} & \makecell[l]{47.61$^{*}_{+}$} \\

Reranker-Prompt$_{hp}$     & \makecell[l]{\textbf{83.88$^{*}_{+}$}}          & \makecell[l]{\textbf{74.98$^{*}_{+}$}}          & \makecell[l]{\textbf{81.99$^{*}_{+}$}}             & \makecell[l]{\textbf{73.35$^{*}_{+}$}}           & \makecell[l]{\textbf{70.03$^{*}_{+}$} }         & \makecell[l]{\textbf{62.21$^{*}_{+}$}}  & \textbf{85.54$^{*}_{+}$} & \textbf{64.37$^{*}_{+}$} & \textbf{58.82$^{*}_{+}$} & \textbf{72.29$^{*}_{+}$} & \textbf{85.28$^{*}_{+}$} & \textbf{90.07$^{*}_{+}$}  & \makecell[l]{\textbf{49.02$^{*}_{+}$}}          & \makecell[l]{\textbf{63.13 $^{*}_{+}$}}          & \makecell[l]{25.94$^{*}_{+}$ }          & \makecell[l]{39.94$^{*}_{+}$}          & \makecell[l]{\textbf{36.64$^{*}_{+}$}}             & \makecell[l]{\textbf{51.58$^{*}_{+}$}}            & \makecell[l]{26.82$^{*}_{+}$}  & \makecell[l]{\textbf{41.44$^{*}_{+}$}} & \makecell[l]{48.21$^{*}_{+}$} \\
\toprule
\end{tabular}
\label{table:out-of-domain reranking}
}}
\subtable[Out-of-domain neural IR pipeline. FT is the traditional fine-tuning paradigm, RP$^{1}_{cp}$ is Retriever-Prompt$_{cp}$ and RP$^{2}_{hp}$ is Reranker-Prompt$_{cp}$.]{
\renewcommand\arraystretch{1.25}

\setlength\tabcolsep{2.5pt}
\renewcommand{\cellset}{\renewcommand{\arraystretch}{1.5}} 

\centering
   \scalebox{0.75}{
\begin{tabular}{llllllllllllllllllllllll}
\toprule
\multirow{2}{*}{Method}  & \multicolumn{2}{c}{Reddit$_{RD}$} & \multicolumn{2}{c}{AQ$_{RD}$} & \multicolumn{2}{c}{Trec$_{QA}$} & \multicolumn{2}{c}{NQ$_{QA}$} & \multicolumn{2}{c}{WQ$_{QA}$} & CW$_{DR}$      & RB04$_{DR}$  & Gov2$_{DR}$ \\ 
                         & P@1           & MRR  & P@1           & MRR    & P@1           & MRR             & P@1           & MRR  & P@1           & MRR  & \multicolumn{3}{c}{NDCG@10}\\ \hline \hline 
\multicolumn{21}{c}{Task-specific model for each specific task} \\
FT$^{1}_{sp}$+FT$^{2}_{sp}$            &55.95 &69.74 & 78.56 & 86.49 & 49.03          & 63.15          & 21.96         & 35.51          & 36.66             & 51.59           & 26.07         & 35.10 & 40.79 \\  \hline
\multicolumn{21}{c}{Multi-task model for all tasks } \\   

FT$^{1}_{mark}$+FT$^{2}_{mark}$              &\makecell[l]{60.33$^{*}$}  &\makecell[l]{74.52$^{*}$}   &\makecell[l]{77.97}  &\makecell[l]{85.83} & \makecell[l]{45.98}          & \makecell[l]{60.89}  & \makecell[l]{24.94$^{*}$}          & \makecell[l]{38.77$^{*}$}          & \makecell[l]{33.15}             & \makecell[l]{48.95}           & \makecell[l]{25.63}                   & \makecell[l]{35.03} & \makecell[l]{43.25$^{*}$}  \\
{HF$^{1}$+HF$^{2}$}              &\makecell[l]{{62.32$^{*}$}}  &\makecell[l]{{76.48$^{*}$}}   &\makecell[l]{{78.95$^{*}$}}  &\makecell[l]{{86.88$^{*}$}} & \makecell[l]{{47.86$^{*}$}}          & \makecell[l]{{61.23$^{*}$}}  & \makecell[l]{{25.39$^{*}$}}          & \makecell[l]{{39.12$^{*}$}}          & \makecell[l]{{35.49$^{*}$}}             & \makecell[l]{{50.37$^{*}$}}           & \makecell[l]{{25.98$^{*}$}}                   & \makecell[l]{{35.57$^{*}$}} & \makecell[l]{{43.72$^{*}$}}  \\ \hdashline
RP$^{1}_{cp}$+RP$^{2}_{hp}$ \    &\textbf{64.78}$^{*}_{+}$ &\textbf{78.47}$^{*}_{+}$   & \textbf{86.39}$^{*}_{+}$ & \textbf{91.03}$^{*}_{+}$ & \makecell[l]{\textbf{53.44}$^{*}_{+}$}          & \makecell[l]{\textbf{67.56}$^{*}_{+}$ }          & \makecell[l]{\textbf{26.03}$^{*}_{+}$ }           & \makecell[l]{\textbf{40.05}$^{*}_{+}$}          & \makecell[l]{\textbf{36.73}$^{*}_{+}$}              & \makecell[l]{\textbf{51.62}$^{*}_{+}$}              & \makecell[l]{\textbf{27.50}$^{*}_{+}$}  & \makecell[l]{\textbf{36.92}$^{*}_{+}$} & \makecell[l]{\textbf{45.33}$^{*}_{+}$} \\

\toprule
\end{tabular}
  \label{table:out-of-domain pipeline}
}
}
\end{table*}

\subsubsection{New Task Adaptation} \label{new task}
{Furthermore, we investigate the few-shot learning capability of the multi-task model when presented with new tasks. Specifically, following the leave-one-out method, for the set of three (or five) specific text matching tasks, we choose two (or four) of them for mixed training, which we refer to as the multi-task${sub}$ model. The remaining task is then designated as the new task for the multi-task${sub}$ model. This allows us to assess the model's ability to capture the essential matching signals shared across tasks in multi-task learning and adapt the learned signals to unseen tasks with only a limited number of examples available. In NIR-Prompt$_{hp}$, the prompt tokens of the new task are obtained by the weighted sum of the prompt tokens of the other four tasks, and the weight of each task is a trainable variable in few-shot learning. The w/o fusion means do not fuse the prompts of other tasks and trains directly from scratch under few-shot settings. In order to better demonstrate the promotion effect of other tasks on low-resource learning, we also use Fine-tuning$_{sp}$ to directly perform few-shot learning on each task. For the new task, we select 32 positive and 32 negative samples for training and observe the performance of the model on the testing set. The experimental results are shown in Table~\ref{few-shot retrieval},~\ref{few-shot reranking} and~\ref{few-shot pipeline}. The models based on multi-task$_{sub}$ get better few-shot learning performance than Fine-tuning$_{sp}$. Besides, Multi-task$_{sub}$ based on NIR-Prompt has better few-shot learning ability in new tasks than that based on other methods. This indicates that the information of other tasks can improve the adaptability of the model to new tasks and NIR-Prompt performs best. In addition, the prompts of other tasks contain descriptions that adapt essential matching signals to tasks, and the integration of them can lead to a good initialization of the description for the new task, which further promotes few-shot learning. The experiment result can answer the third research question that the multi-task learning in the essential matching module efficiently learns the essential matching signals that can shared across tasks and the learned signals can facilitate the new task adaptation in low-resource scenarios.}

\begin{table*}[htbp]
  \caption{New-task adaptation performance of the models.
  Each task in this table is a new task not included in mixed datasets for multi-task$_{sub}$ (leave-one-out). Results with significant performance improvement with p-value $ \leq 0.05$ compared with all multi-task models of baselines are denoted as `$+$'.}
\subtable[New-task adaptation retrieval]{
\renewcommand\arraystretch{1.15}

\setlength\tabcolsep{11pt}
\renewcommand{\cellset}{\renewcommand{\arraystretch}{0.5}} 

\centering
   \scalebox{0.9}{
\begin{tabular}{llllllllll}
\toprule
\multirow{2}{*}{Method} & \multicolumn{2}{c}{Trivia$_{QA}$} & \multicolumn{2}{c}{DailyDialog$_{RD}$} & MQ2007$_{DR}$ \\ 
                        &  P@1(\%)           & MRR(\%)             & P@1(\%)           & MRR(\%)         & NDCG@10(\%) \\ \hline  \hline 
\multicolumn{6}{c}{Few-shot learning on each task} \\
Fine-tuning$_{sp}$    & {12.91} & {20.47} & {23.44} & {29.17}  & {9.23} \\       \hline
\multicolumn{6}{c}{Few-shot learning based on multi-task$_{sub}$ on each task} \\
Fine-tuning$^{1}_{multi}$   & {27.08} & {35.23}  & {32.56}  & {38.44}    & {10.29}\\

Fine-tuning$^{1}_{mark}$   & {26.73} & {34.91}   & {31.15}  & {37.05}    & {10.12} \\
{ATTEMPT$^{1}$} & {{31.58}} & {{39.85}}   & {{35.27}}  & {41.10}    & {{11.79}} \\
{HyperFormer$^{1}$} & {{31.62}} & {{39.90}}   & {{36.03}}  & {{41.72}}    & {{11.50}} \\
{AdapterFusion$^{1}$} & {{30.43}} & {{39.02}}   & {{35.11}}  & {{40.98}}    & {{11.42}} \\
\hdashline
Retriever-Prompt$_{cp}$    & {\textbf{32.45$_{+}$}} & {\textbf{40.72$_{+}$}} & {\textbf{37.89$_{+}$}}   & {\textbf{43.01$_{+}$}}   & {\textbf{12.47$_{+}$}} \\ 
- w/o fusion   & {28.58}   & {36.33}      & {34.05}      & {40.12}   & {11.34} \\
\toprule
\end{tabular}
\label{few-shot retrieval}
}
}

\subtable[New-task adaptation reranking]{
\renewcommand\arraystretch{1.15}

\setlength\tabcolsep{2pt}
\renewcommand{\cellset}{\renewcommand{\arraystretch}{0.5}} 

\centering
   \scalebox{0.85}{
\begin{tabular}{llllllllll}
\toprule
\multirow{2}{*}{Method} & \multicolumn{2}{c}{Trivia$_{QA}$} & \multicolumn{2}{c}{DailyDialog$_{RD}$} & \multicolumn{2}{c}{MSRP+QQP$_{PI}$}  & \multicolumn{2}{c}{MultiNLI$_{NLI}$} & MQ2007$_{DR}$ \\ 
                        &  P@1(\%)           & MRR(\%)             & P@1(\%)           & MRR(\%)               & Acc.(\%)           & F$_1$(\%)             & Acc.(\%)           & F$_1$(\%)            & NDCG@10(\%) \\ \hline  \hline 
\multicolumn{10}{c}{Few-shot learning on each task} \\
Fine-tuning$^{2}_{sp}$  & {18.78}  & {24.13}  & {32.83} & {38.97}  & {64.93} & {60.87} & {35.76} & {28.70} & {33.87} \\       \hline
\multicolumn{10}{c}{Few-shot learning based on multi-task$_{sub}$ on each task} \\
MTL$_{T5}$   & {29.25}  & {35.85} & {39.65}  & {45.92}  & {71.13}   & {67.93}   & {48.78}  & {40.39}         & {35.34}\\
MT-DNN   & {31.68} & {37.39} & {44.69}    & {50.97}     & {68.17}  & {65.49}    & {59.16}  & {52.12}         & {35.12} \\ 
Fine-tuning$^{2}_{multi}$  & {34.35}   & {40.11}  & {45.18}   & {51.03}   & {67.01} & {64.35}             & {52.50}   & {45.98}     & {36.11}\\

Fine-tuning$^{2}_{mark}$  & {31.73}   & {37.49}  & {44.07}    & {50.06}      & {70.90}  & {67.92}       & {59.98}   & {53.45}  & {36.02}\\
{ATTEMPT$^{2}$}  & {38.03}   & {43.72}   & {45.91} & {51.41}  & {73.52}  & {70.67}       & {60.71}   & {54.75}  & {36.18}\\
{HyperFormer$^{2}$}  & {38.15}   & {43.80} & {46.03} & {51.50}  & {73.74}  & {70.82}       & \textbf{{62.09}}   & \textbf{{56.81}}  & {36.34}\\
{AdapterFusion$^{2}$}  & {37.21}   & {43.05} & {45.58} & {51.23}     & {74.01} & {71.10}       & {61.98}   & {56.67} & {36.21}\\
\hdashline

Reranker-Prompt$_{hp}$  & {\textbf{39.85$_{+}$}}     & {\textbf{45.13$_{+}$}} & {\textbf{46.49$_{+}$}}   & {\textbf{52.59$_{+}$}}      & {\textbf{74.15$_{+}$}}    & {\textbf{71.33$_{+}$}}        & 62.02   & 56.73      & {\textbf{37.18$_{+}$}} \\ 
- w/o fusion   & {36.87}   & {42.37}  & {45.02}    & {51.09}   & {71.01}     & {68.09}      & {60.92}     & {54.33}            & {36.28} \\
\toprule
\end{tabular}
\label{few-shot reranking}
}
}
\subtable[New-task adaptation neural IR pipeline. FT is the traditional fine-tuning paradigm, RP$^{1}_{cp}$ is Retriever-Prompt$_{cp}$ and RP$^{2}_{hp}$ is Reranker-Prompt$_{cp}$.]{
\renewcommand\arraystretch{1.25}

\setlength\tabcolsep{12pt}
\renewcommand{\cellset}{\renewcommand{\arraystretch}{0.5}} 

\centering
   \scalebox{0.9}{
\begin{tabular}{llllllllll}
\toprule
\multirow{2}{*}{Method} & \multicolumn{2}{c}{Trivia$_{QA}$} & \multicolumn{2}{c}{DailyDialog$_{RD}$} & MQ2007$_{DR}$ \\ 
                        &  P@1(\%)           & MRR(\%)             & P@1(\%)           & MRR(\%)         & NDCG@10(\%) \\ \hline  \hline
\multicolumn{6}{c}{Few-shot learning on each task} \\
FT$^{1}_{sp}$+FT$^{2}_{sp}$  & {15.45}   & {22.68} & {30.36} & {35.42} & {15.43} \\       \hline
\multicolumn{6}{c}{Few-shot learning based on multi-task$_{sub}$ on each task} \\

FT$^{1}_{mark}$+FT$^{2}_{mark}$ & {27.92}   & {36.48}   & {39.75} & {45.36}            & {18.55}\\
{HF$^{1}$+HF$^{2}$} & {31.21}   & {40.42}   & {46.35} & {52.10} & {20.01}\\
\hdashline
RP$^{1}_{cp}$+RP$^{2}_{hp}$  & {\textbf{33.49$_{+}$}}   & {\textbf{42.83$_{+}$}}    & {\textbf{47.61$_{+}$}}      & {\textbf{53.04$_{+}$}}     & {\textbf{20.85$_{+}$}} \\ 
\toprule
\end{tabular}
\label{few-shot pipeline}
}
}
\end{table*}

\subsubsection{Performance on BEIR}
We also evaluate the performance of NIR-Prompt and the traditional fine-tuning paradigm on BEIR, a heterogeneous benchmark for testing the generalization ability of retrieval models. In this experiment, we test the generalization performance of dense retrieval, reranking, and the neural information retrieval pipeline on BEIR respectively. The models used for evaluation are trained on Table~\ref{table:basic-datasets retrieval} for dense retrieval and Table~\ref{table:basic-datasets reranking} for reranking, which are consistent with Section~\ref{in-domain} and \ref{out-of-domain}. The result shown in Table~\ref{table:beir} indicates that NIR-Prompt shows better performance than the traditional fine-tuning paradigm in dense retrieval, reranking, and neural information retrieval pipeline, which further demonstrates the generalization ability of our method.
\begin{table*}[t]
  \caption{Result on BEIR. FT is the traditional fine-tuning paradigm, RP$^{1}_{cp}$ is Retriever-Prompt$_{cp}$ and RP$^{2}_{hp}$ is Reranker-Prompt$_{cp}$. \textbf{Boldface} indicates the best results for each stage. Results with significant performance improvement with p-value $ \leq 0.05$ compared with traditional fine-tuning baselines are denoted as `$+$'. {The evaluation metric is NDCG@10.}}
  \label{table:cross-datasets-results retrieval}
\renewcommand\arraystretch{1.25}
\setlength\tabcolsep{1.25pt}%
\renewcommand{\cellset}{\renewcommand{\arraystretch}{1.5}} 

\centering
   \scalebox{0.9}{
    \begin{tabular}{@{}l|ccc|c@{\hspace{15pt}}cc|cc}
    \toprule
	     & FT$^{1}_{multi}$ & FT$^{1}_{mark}$ & RP$^{1}_{cp}$ & FT$^{2}_{multi}$ & FT$^{2}_{mark}$ & RP$^{2}_{hp}$ & FT$^{1}_{mark}$+FT$^{2}_{mark}$ & RP$^{1}_{cp}$+RP$^{2}_{hp}$      \\
        & \multicolumn{3}{c|}{Dense Retrieval} & \multicolumn{3}{c|}{Reranking (BM25 retrieved)} & \multicolumn{2}{c}{Neural IR Pipeline}\\
	 \hline
	 TREC-COVID   & 45.53 & 46.64 & \textbf{49.31$_{+}$} & 54.61 & 54.98 & \textbf{57.87$_{+}$} & 50.37 & \textbf{55.77$_{+}$}\\
	 Touch\'e     & 16.56 & 17.51 & \textbf{21.92$_{+}$}  & 22.02 & 22.53 & \textbf{23.92$_{+}$} & 19.75 & \textbf{23.01$_{+}$}\\
	 DBPedia     & 9.13 & 9.26 & \textbf{11.32$_{+}$} & 19.43 & 19.56 & \textbf{21.43$_{+}$} & 15.71 & \textbf{17.92$_{+}$} \\
	 NFCorpus    & 18.63  & 18.82 & \textbf{21.27$_{+}$} & 28.45  & 29.03 & \textbf{32.25$_{+}$} & 24.79 & \textbf{28.26$_{+}$}\\
	 Quora       & 24.19 & 24.51 & \textbf{55.41$_{+}$}  &85.44 & 85.68 & \textbf{87.29$_{+}$} & 69.78 & \textbf{82.33$_{+}$}\\
	 HotpotQA    & 38.57 & 39.42 & \textbf{41.70$_{+}$} & 47.22 & 47.69 & \textbf{50.43$_{+}$}  & 44.64 & \textbf{48.39$_{+}$} \\
	 FEVER       & 55.32 & 56.07 & \textbf{59.87$_{+}$} & 65.43 & 65.71 & \textbf{68.19$_{+}$} & 63.56 & \textbf{67.03$_{+}$} \\
	 FiQA        & 12.01 & 12.28 & \textbf{14.54$_{+}$} & 25.58  & 25.61 & \textbf{27.05$_{+}$} & 20.70 & \textbf{23.53$_{+}$} \\
     ArguAna     & 21.32 & 22.30 & \textbf{22.61$_{+}$} & 23.41 & 23.55 & \textbf{25.43$_{+}$} & 22.04 & \textbf{23.45$_{+}$}\\
     NQ          & 43.92 & 44.31 & \textbf{45.69$_{+}$}  & 44.03 & 44.59 & \textbf{45.65$_{+}$} & 46.13 & \textbf{47.80$_{+}$} \\
	 SciFact     &20.65  &21.43  &\textbf{41.46$_{+}$}  & 44.96 & 45.32 & \textbf{54.97$_{+}$} & 30.25 & \textbf{50.88$_{+}$}\\
	 SCIDOCS     &8.59  &9.01  &\textbf{10.35$_{+}$}  & 12.31 & 12.55 & \textbf{13.75$_{+}$} & 10.36 & \textbf{12.09$_{+}$} \\
   Climate-FEVER  &10.01  &10.13  &\textbf{15.30$_{+}$}  & 13.48 & 13.66 & \textbf{17.42$_{+}$} & 11.93 & \textbf{16.37$_{+}$} \\
	 CQADupStack &16.20  & 16.31 & \textbf{17.29$_{+}$} & 26.38 & 26.49 & \textbf{27.33$_{+}$} & 22.35 & \textbf{24.02$_{+}$} \\
    \toprule
    \end{tabular} }
    \label{table:beir}
\end{table*}

\subsection{Model Analysis}
The quality of learned prompt tokens is very important for the multi-task generalization of text matching models. We first measure the specificity of the information stored in the prompt tokens by evaluating on handcrafted `QA vs. PI' task. Then, we explore the connections between task prompt tokens and their composability.
\subsubsection{Ability to Distinguish Tasks}
We construct handcrafted `QA vs PI' datasets to verify it and the examples come from given QA datasets (Trec, WQ, and NQ). For each question, we provide it with two candidate texts, one is the passage containing the answer (a positive sample in original QA datasets), denoted as $a$, and one is the question itself, denoted as $q$. Since $q$ is exactly the same as the question, while can not be the answer to the question, so it can only reflect the matching relation in PI task. In contrast, $a$ contains the answer to the question and can be used to denote the matching relation in QA task. Due to the extra answer information in $a$, its matching score under PI task will be lower than $q$ which is exactly the same as the question. Therefore, by altering task marks between QA and PI and comparing the matching scores of two candidates, we can evaluate the ability of the multi-task model to distinguish tasks.
Table~\ref{table:dis} indicates that NIR-Prompt can improve the ability of the multi-task model to distinguish different tasks. Even though other multi-task learning methods distinguish tasks by different means and are trained on datasets of multiple tasks, they still tend to the exact matching signals but cannot distinguish tasks well. This is also the key reason why NIR-Prompt works better.
\begin{table}[t]
  \caption{Illustrate the ability to distinguish tasks.}
  \label{table:dis}
\renewcommand\arraystretch{1.2}
\setlength\tabcolsep{11.2pt}
\renewcommand{\cellset}{\renewcommand{\arraystretch}{0.5}} 
\scalebox{0.9}{
\begin{tabular}{lllllll}
\toprule
\multirow{3}{*}{Method} & \multicolumn{3}{c}{As QA Task} & \multicolumn{3}{c}{As PI Task} \\
                        & Trec& WQ& NQ & Trec & WQ& NQ\\
                        & P$_{a>q}$     & P$_{a>q}$              & P$_{a>q}$ & P$_{q>a}$          & P$_{q>a}$   & P$_{q>a}$ \\ \hline 
MTL$_{T5}$             & 38.35         & 6.39          & 9.03 & 76.07         & 97.09          & 95.45 \\
MT-DNN             & 27.76         & 29.11          & 19.94 & 98.67         & 99.56          & 99.70 \\ 
Fine-tuning$_{multi}^{2}$             & 20.91         & 13.47          & 4.86 & 79.09         & 86.53          & 95.14 \\
Fine-tuning$_{mark}^{2}$             & 40.75         & 34.82          & 10.49 & 63.70         & 72.06          & 97.37 \\ 
{ATTEMPT$^{2}$}       & {60.42}         & {61.79}  & {50.71}      & {99.01}   & {99.71}   & {99.67} \\ 
{HyperFormer$^{2}$}   & {62.37}        & {61.88}  &  {55.63}     & {98.79}   & {99.10}   & {99.20} \\ 
{AdapterFusion$^{2}$} & {60.59}         & {61.03}  &  {53.01}     & {99.15}   & {99.22}   & {99.31} \\ 
\hdashline
Reranker-Prompt$_{hp}^{2}$  & \textbf{71.98}         & \textbf{70.68}          & \textbf{59.92}    & \textbf{100.00}         & \textbf{100.00} & \textbf{100.00}     \\  
\toprule
\end{tabular}
}
\end{table}

\subsubsection{Relationships between Tasks}
\begin{figure}[t]
    \centering
    \subfigure[{Distribution of the cosine similarity for prompt tokens}]{
	\includegraphics[width=2.5in]{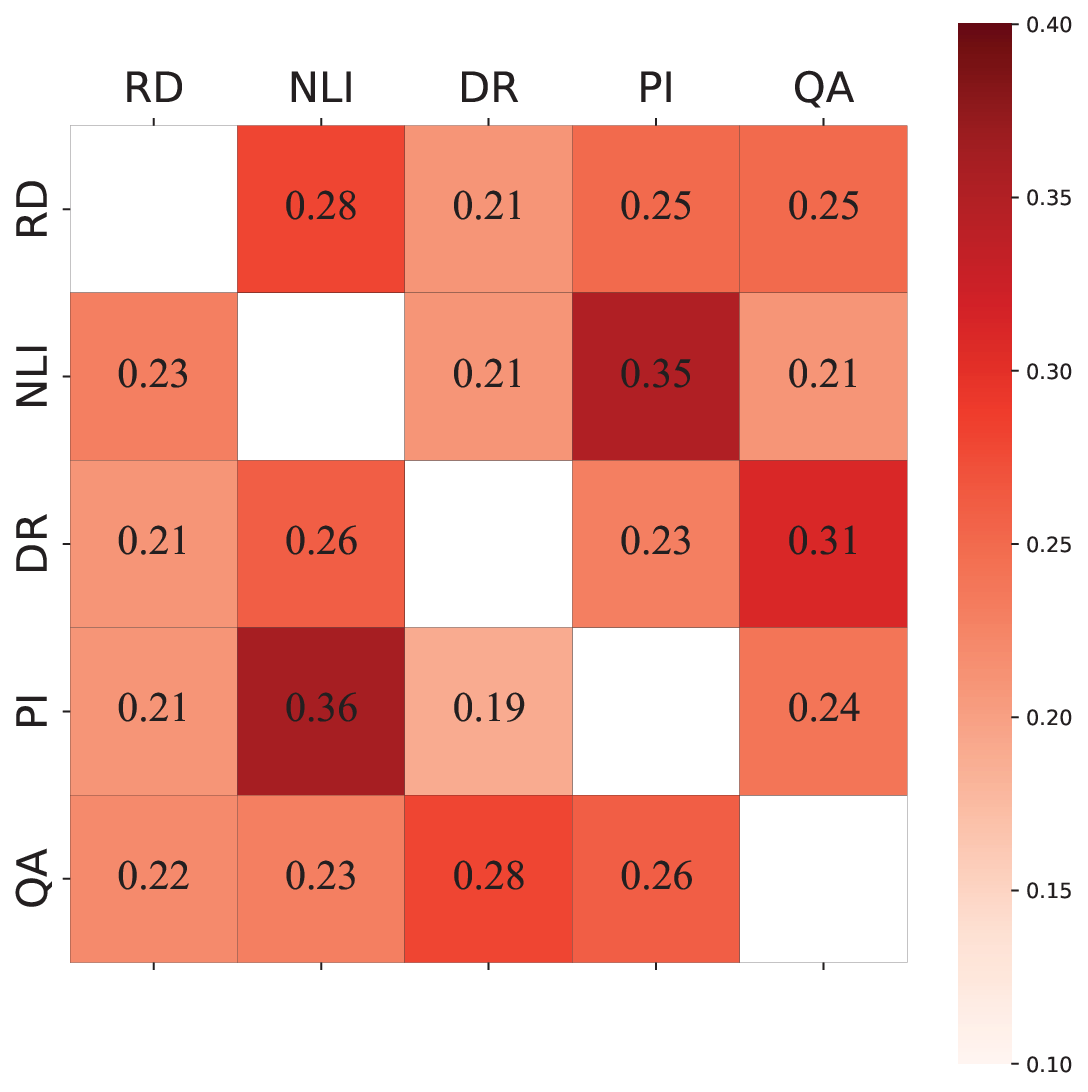}
	\label{prompt heatmap}
    }
    \subfigure[{Distribution of the weights for new task prompt tokens}]{
        \includegraphics[width=2.5in]{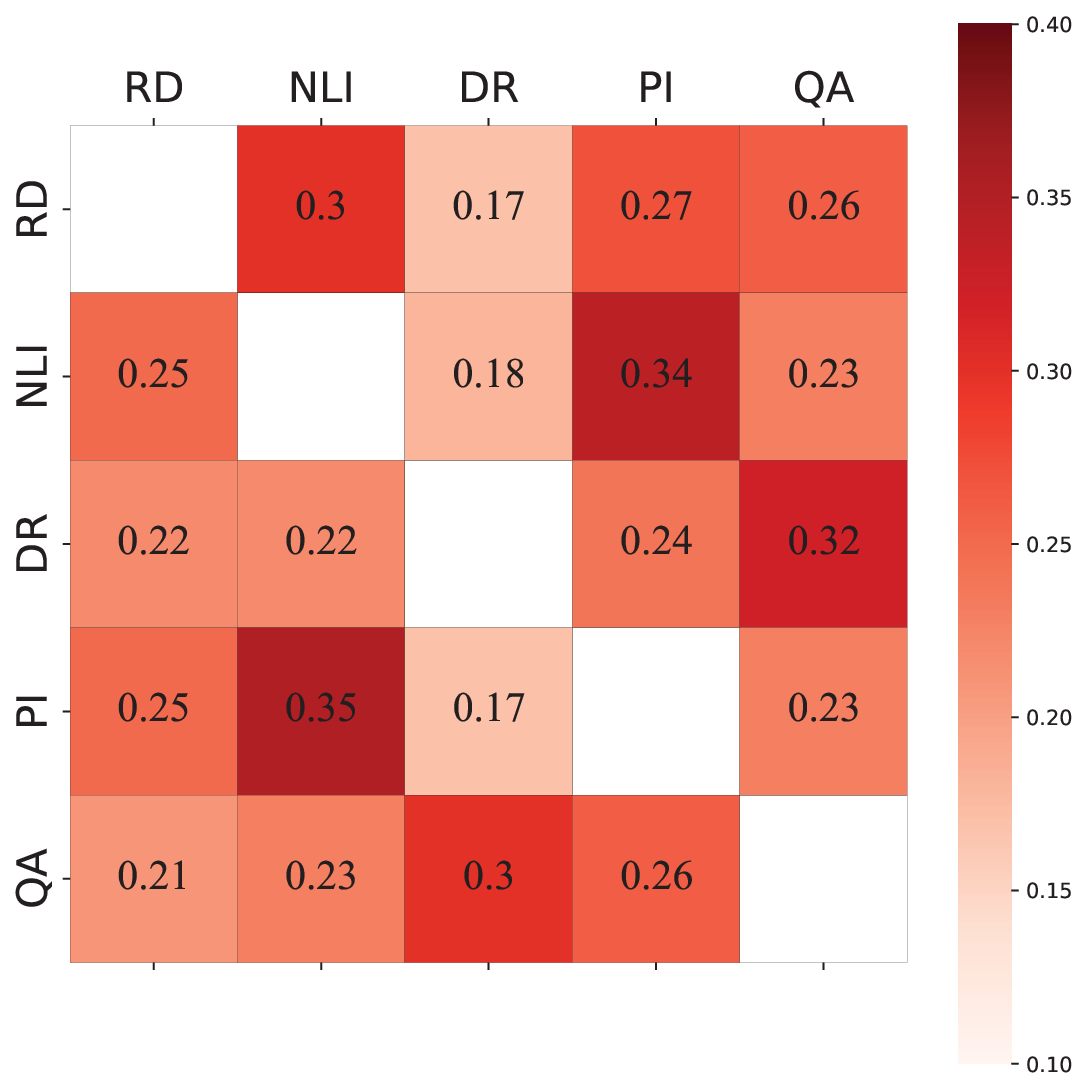}
    \label{weight}
    }
    \caption{{Task relationships shown by the prompt tokens. The sum of each row is 1, and the value is rounded to two decimal places. The value in (a) represents the distribution of cosine similarity between tasks in each row and the other tasks. The value in (b) represents the distribution of the fusion weights of other tasks for each new task in row (obtained by Section~\ref{new task}). There is no similarity calculation between the same tasks, so the diagonal elements are zero.}}
    \label{heatmap}
\end{figure}

We explore the relationships between tasks using the cosine similarity of task prompt token embeddings learned in Section~\ref{in-domain} and its heatmap is shown in Figure~\ref{prompt heatmap}. We can see that NLI is similar to PI because they both focus on exact matching signals, and DR is similar to QA because both exact and semantic matching signals are important in these tasks. The similarity between different tasks is generally consistent with our prior knowledge of the task. The heatmap of the fusion weights of tasks for each new task obtained in Section~\ref{new task} is shown in Figure~\ref{weight}. This weight distribution is consistent with the embeddings similarity distribution of the prompt tokens between tasks, which further supports the rationality of fusing different tasks for new task adaptation.

\subsubsection{{Compatibility to SOTA IR Models}}
{
Our method is a general training framework that can be combined with existing IR models to further improve their performance in generalization. In this section, we explore the effectiveness of our method based on RetroMAE~\cite{retromae}, a state-of-the-art IR model on both in-domain and out-of-domain generalization that has been pre-trained on the large self-supervised corpus. Specifically, we use the traditional fine-tuning method and our method (Retriever-Prompt) to train RetroMAE on mixed datasets in Table~\ref{table:basic-datasets retrieval} respectively and compare their in-domain and out-of-domain generalization ability on the datasets in Table~\ref{table:basic-datasets retrieval} and~\ref{table:datasets}. The experimental results are shown in Table~\ref{retromae_exp}. Compared with the traditional fine-tuning method, our method can further improve the in-domain and out-of-domain ability of RetroMAE.}

\begin{table}[h]
  \caption{{In-domain and out-of-domain generalization ability comparision between RetroMAE trained by traditional fine-tuning method (RetroMAE$_{FT}$) and Retriever-Prompt (RetroMAE$_{RP}$). \textbf{Boldface} indicates the best results of multi-task models and the results over RetroMAE$_{sp}$ are denoted as `$*$'. Results with significant performance improvement with p-value $ \leq 0.05$ compared with RetroMAE$_{FT}$ are denoted as `$+$'.}}
  \label{retromae_exp}
  \subtable[In-domain multi-task]{
\renewcommand\arraystretch{1.1}
\setlength\tabcolsep{15pt}
\renewcommand{\cellset}{\renewcommand{\arraystretch}{0.5}} 
   \scalebox{0.9}{
\begin{tabular}{llllllllll}
\toprule
\multirow{2}{*}{Method} & \multicolumn{2}{c}{Trivia$_{QA}$} & \multicolumn{2}{c}{DailyDialog$_{RD}$} & MQ2007$_{DR}$ \\ 
                        &  P@1(\%)           & MRR(\%)             & P@1(\%)           & MRR(\%)         & NDCG@10(\%) \\ \hline  \hline 
\multicolumn{10}{c}{Task-specific model for each specific task} \\    
{RetroMAE$_{sp}$}  & {58.91}          & {68.40}          & {63.19}            & {68.97}            & {36.08}  \\  \hline  

{RetroMAE$_{FT}$}      & \makecell[l]{{59.20$^{*}$}}          & \makecell[l]{{68.61$^{*}$}}          & \makecell[l]{{61.29}}             & \makecell[l]{{67.10}}             & {34.75} \\

{RetroMAE$_{RP}$}     & \makecell[l]{{\textbf{61.47$^{*}_{+}$}}}          & \makecell[l]{{\textbf{71.15$^{*}_{+}$}}}          & \makecell[l]{{\textbf{64.22$^{*}_{+}$}}}             & \makecell[l]{{\textbf{69.98$^{*}_{+}$}}}             & {\textbf{36.75$^{*}_{+}$}}      \\
\toprule
\end{tabular}
}
    \label{retromae_id}
}
\subtable[Out-of-domain generalization]{
\renewcommand\arraystretch{1}
\setlength\tabcolsep{3pt}
\renewcommand{\cellset}{\renewcommand{\arraystretch}{0.5}} 
   \scalebox{0.75}{
\begin{tabular}{llllllllllllllllllllllll}
\toprule
\multirow{2}{*}{Method}  & \multicolumn{2}{c}{Reddit$_{RD}$} & \multicolumn{2}{c}{AQ$_{RD}$} & \multicolumn{2}{c}{Trec$_{QA}$} & \multicolumn{2}{c}{NQ$_{QA}$} & \multicolumn{2}{c}{WQ$_{QA}$} & CW$_{DR}$      & RB04$_{DR}$  & Gov2$_{DR}$ \\
                         & P@1           & MRR  & P@1           & MRR    & P@1           & MRR             & P@1           & MRR  & P@1           & MRR  & \multicolumn{3}{c}{NDCG@10}\\ \hline \hline 
\multicolumn{21}{c}{Task-specific model for each specific task} \\
{RetroMAE$_{sp}$}            & {43.56}  & {59.71}  & {86.45} & {92.69} &   {45.91}   & {61.03} & {39.47}  & {50.98}    & {42.66}      & {55.10}    & {30.35}        & {43.22} & {47.54} \\  \hline
\multicolumn{21}{c}{Multi-task model for all tasks } \\   

{RetroMAE$_{FT}$}   &\makecell[l]{{43.09}}   &\makecell[l]{{59.15}}  &\makecell[l]{{86.32}} & \makecell[l]{{92.56}}          & \makecell[l]{{52.47}}  & \makecell[l]{{68.19}}                  & \makecell[l]{{45.21}}           & \makecell[l]{{57.08}}       & \makecell[l]{{42.05}}          & \makecell[l]{{54.87}}                & \makecell[l]{{30.27}} & \makecell[l]{{43.98}}  & \makecell[l]{{47.37}}\\

{RetroMAE$_{RP}$}    &\makecell[l]{{\textbf{43.22$_{+}$}}}   &\makecell[l]{{\textbf{59.37$_{+}$}}}  &\makecell[l]{{\textbf{86.59$^{*}_{+}$}}} & \makecell[l]{{\textbf{92.77$^{*}_{+}$}}}          & \makecell[l]{{\textbf{55.78$^{*}_{+}$}}}  & \makecell[l]{{\textbf{71.59$^{*}_{+}$}}}               & \makecell[l]{{\textbf{47.36$^{*}_{+}$}}}           & \makecell[l]{{\textbf{59.62$^{*}_{+}$}}}       & \makecell[l]{{\textbf{45.39$^{*}_{+}$}}}          & \makecell[l]{{\textbf{57.76$^{*}_{+}$}}}                  & \makecell[l]{{\textbf{30.45$^{*}_{+}$}}} & \makecell[l]{{\textbf{44.52$^{*}_{+}$}}}  & \makecell[l]{{\textbf{49.03$^{*}_{+}$}}}\\

\toprule
\end{tabular}

}
  \label{retromae_ood}
}
\end{table}

\subsubsection{{Analysis of Fixing the Prompt Tokens}}

\begin{figure}[h]
    \centering
    \subfigure[{MRR for QA}]{
	\includegraphics[width=2.6in]{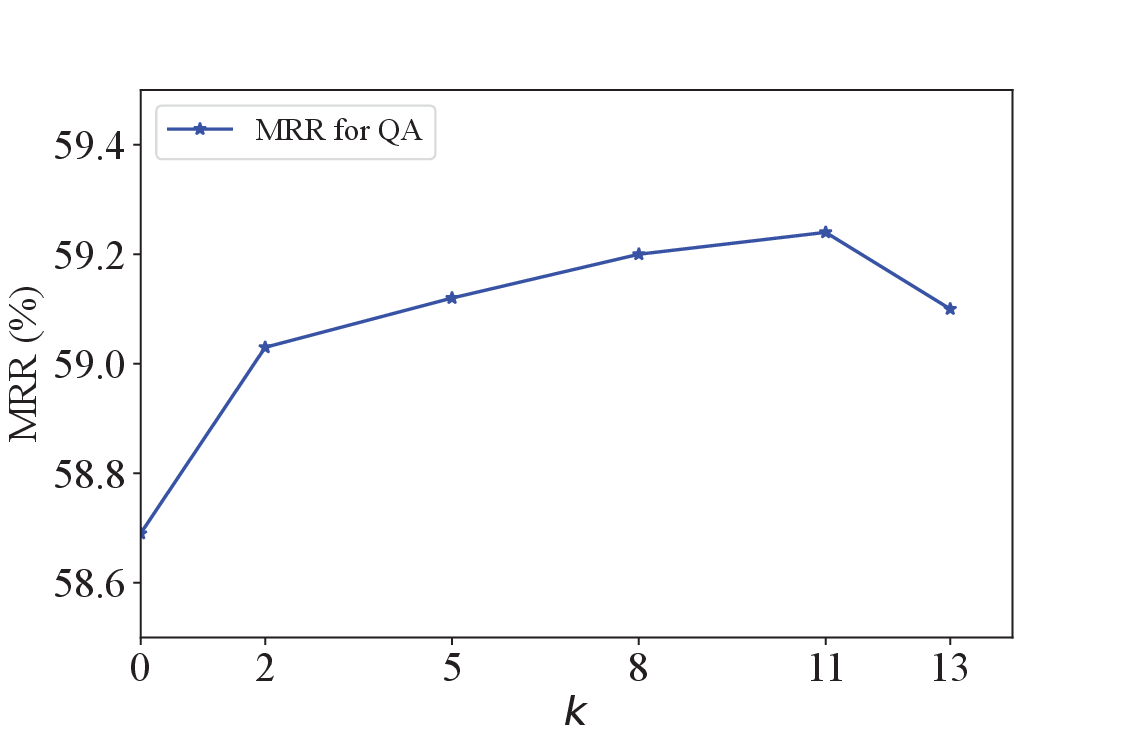}
	\label{k_qa}
    }
        \subfigure[{P@1 for QA}]{
	\includegraphics[width=2.6in]{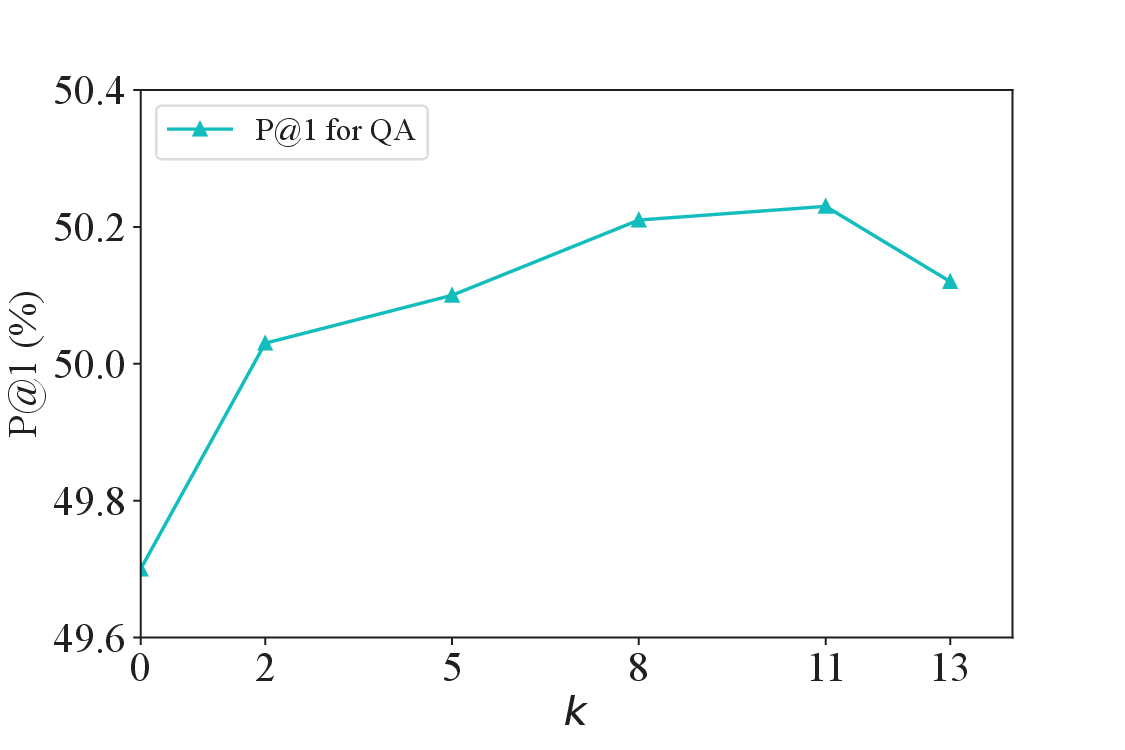}
	\label{k_dia}
    }
        \subfigure[{MRR for Dia}]{
	\includegraphics[width=2.6in]{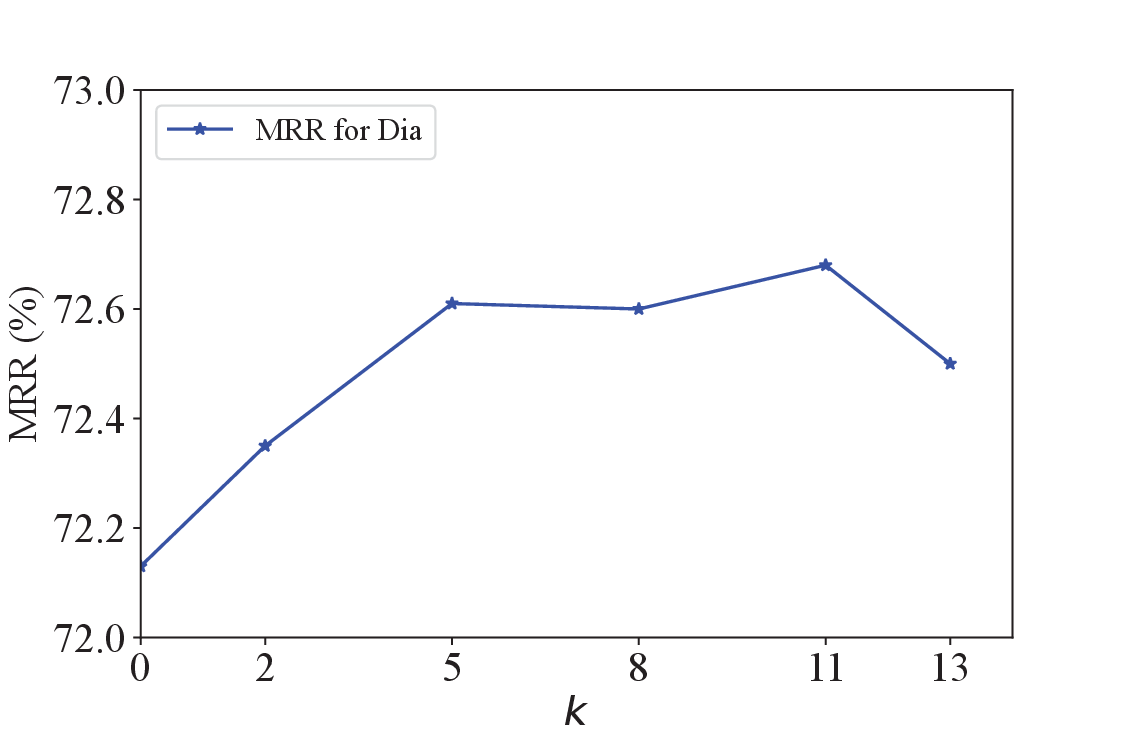}
	\label{k_qa}
    }
        \subfigure[{P@1 for Dia}]{
	\includegraphics[width=2.6in]{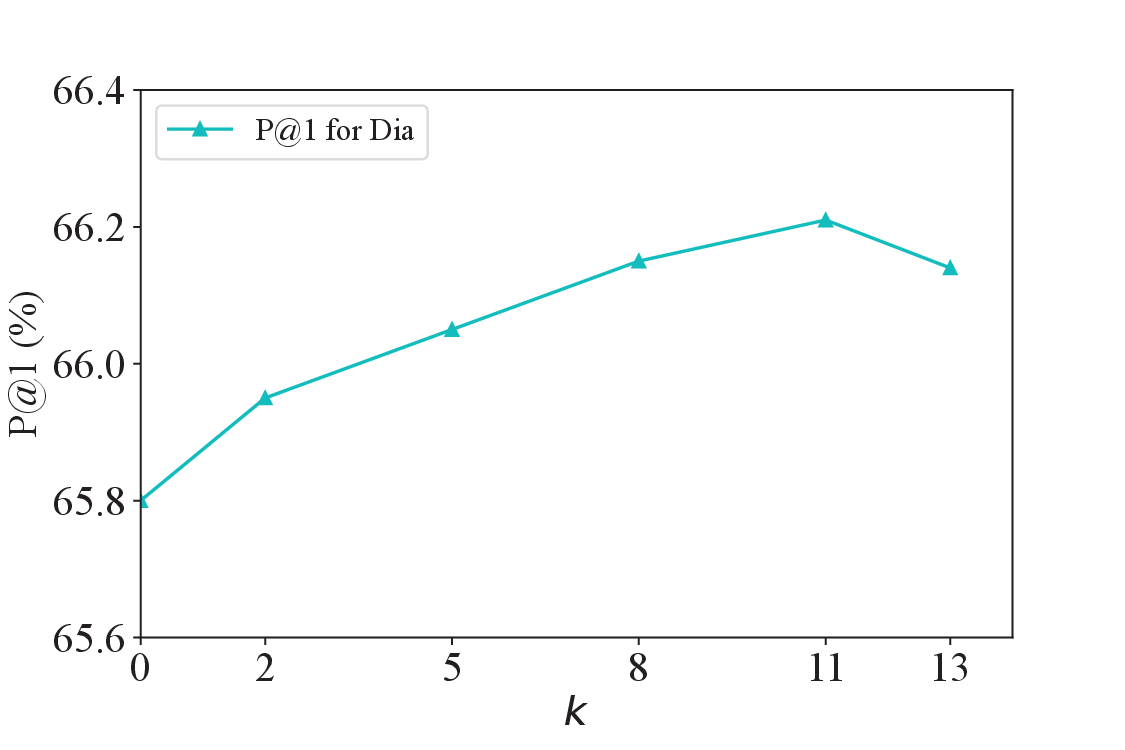}
	\label{k_dia}
    }
    
        \subfigure[{NDCG@10 for DR}]{
	\includegraphics[width=2.6in]{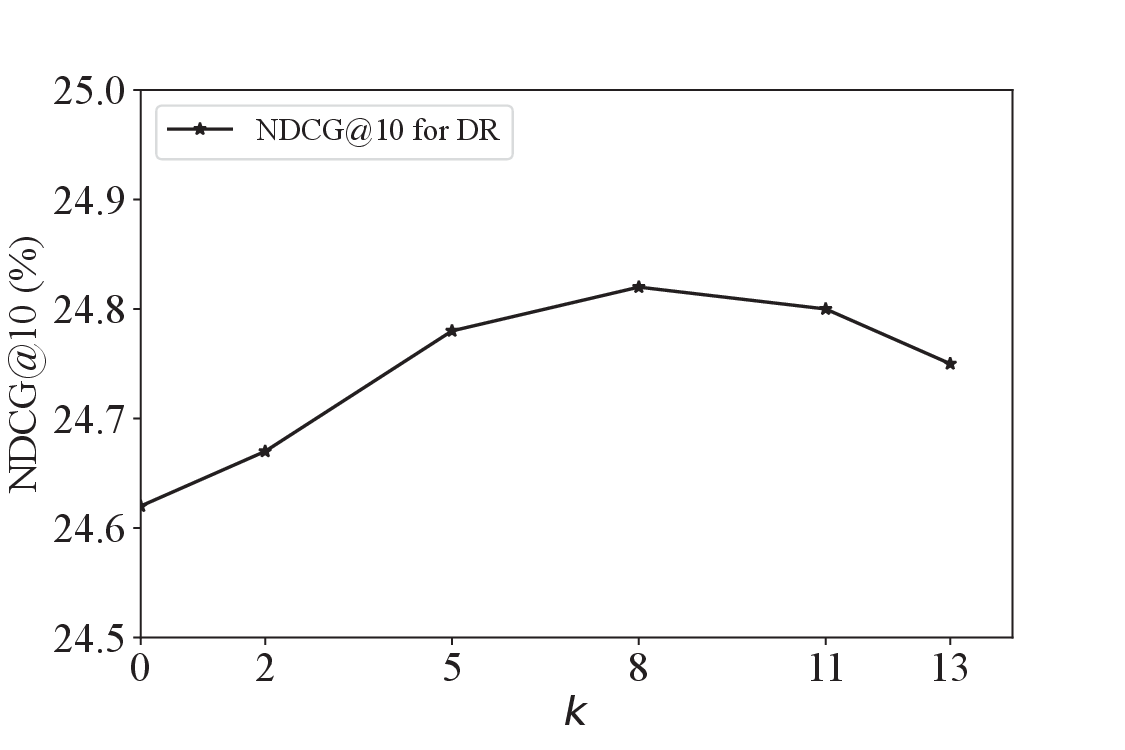}
	\label{k_adhoc}
    }
    \caption{{Performance on various tasks varies with $k$ in retrieval.}}
    \label{K_analysis_retrieval}
\end{figure}

\begin{figure}
    \centering
    \subfigure[{Acc and F1 for PI}]{
	\includegraphics[width=2.6in]{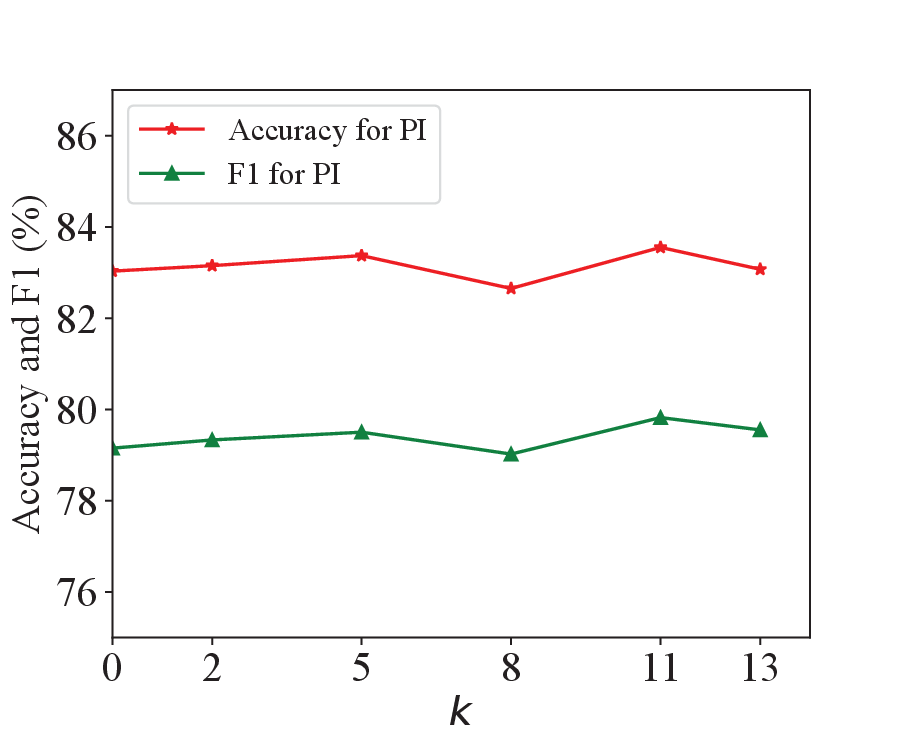}
	\label{k_pi}
    }
    \subfigure[{Acc and F1 for NLI}]{
	\includegraphics[width=2.6in]{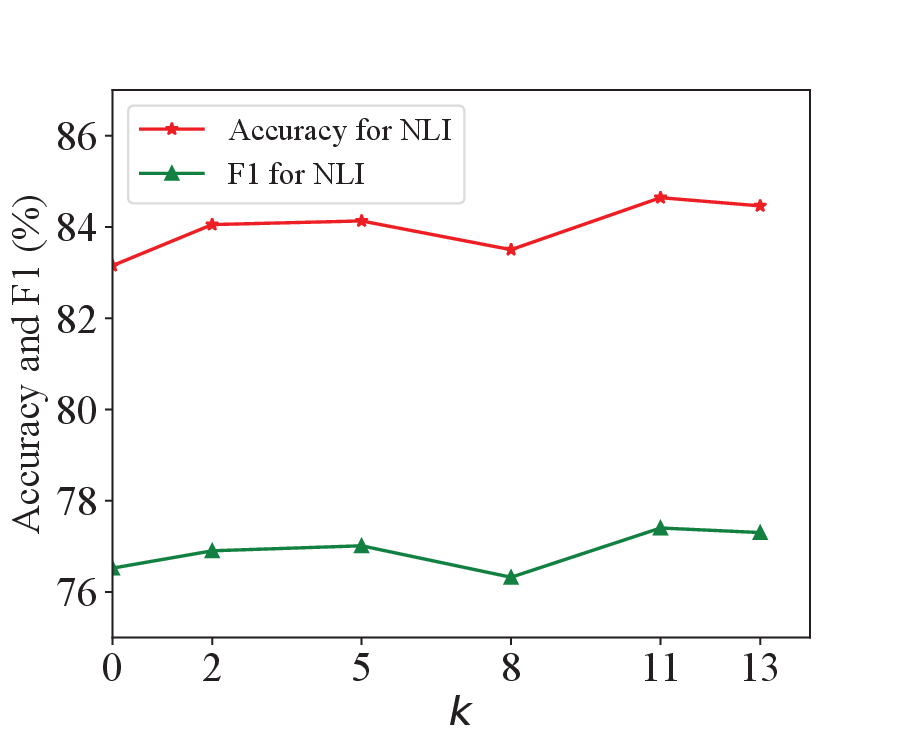}
	\label{k_nli}
    }
    \subfigure[{MRR and P@1 for QA}]{
	\includegraphics[width=2.6in]{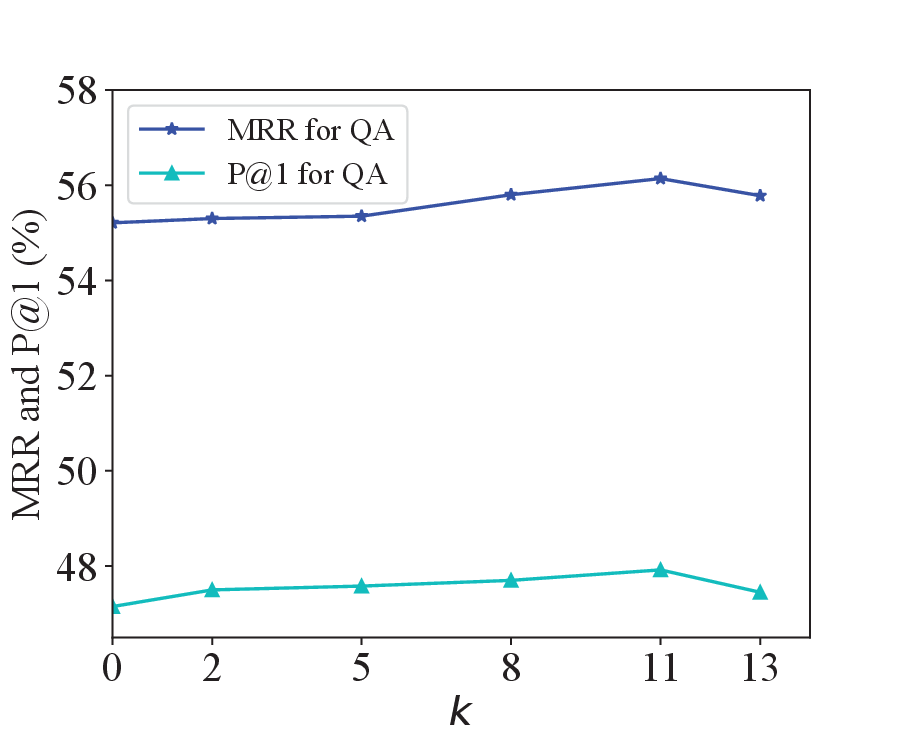}
	\label{k_qa}
    }
        \subfigure[{MRR and P@1 for Dia}]{
	\includegraphics[width=2.6in]{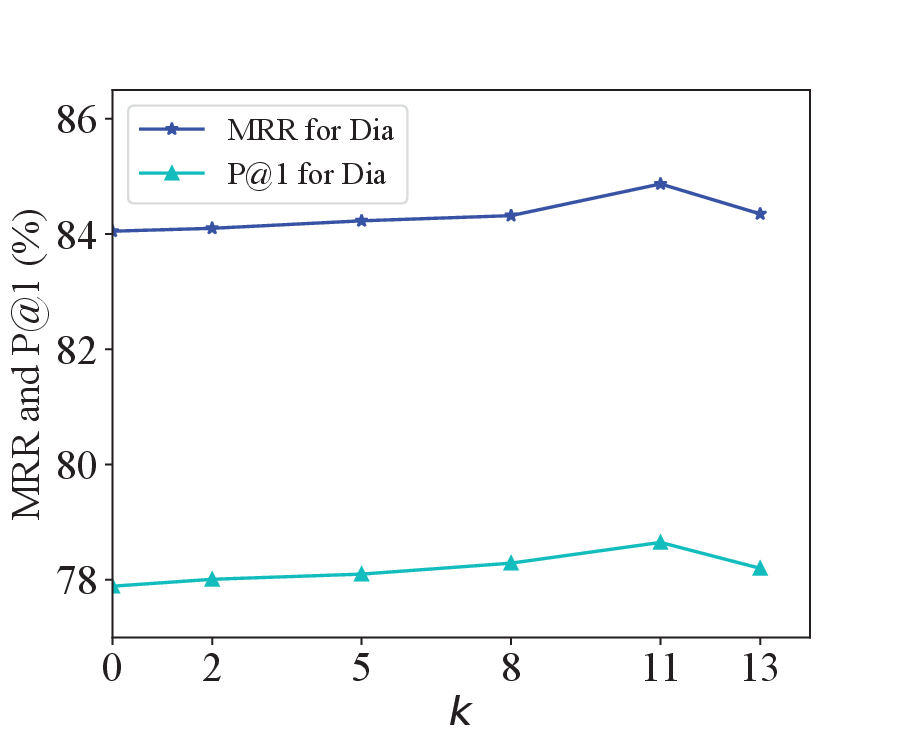}
	\label{k_dia}
    }
        \subfigure[{NDCG@10 for DR}]{
	\includegraphics[width=2.6in]{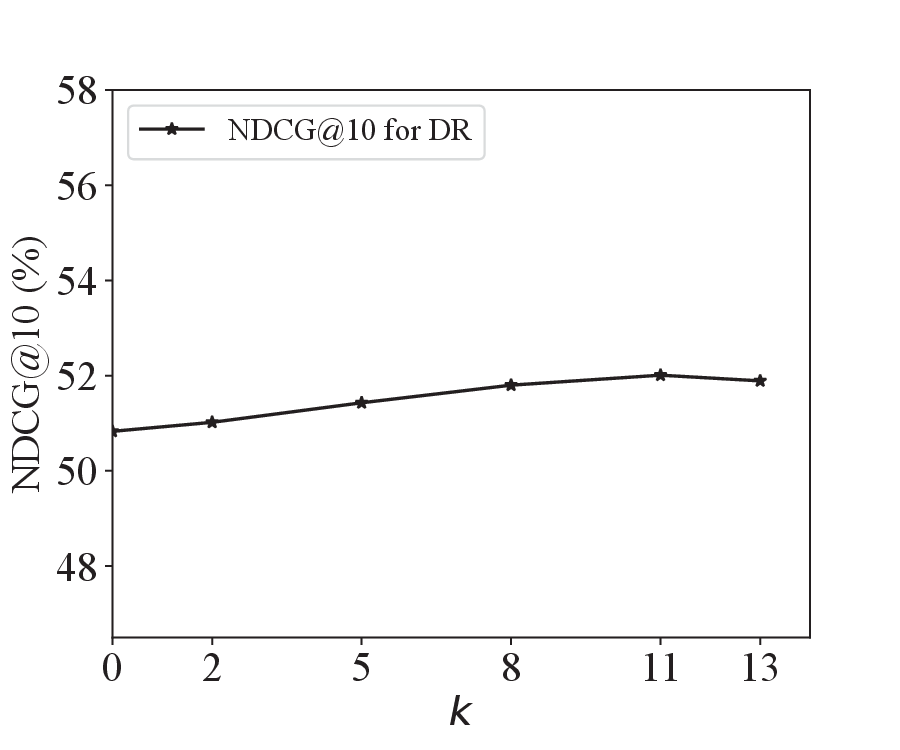}
	\label{k_adhoc}
    }
    \caption{{Performance on various tasks varies with $k$ in reranking.}}
    \label{K_analysis}
\end{figure}

{In Section~\ref{cp retrieval}, we propose that in the training of continuous prompt tokens, we let the embedding of trainable prompt tokens remain fixed in layers $1$ to $k$ in the self-attention. The intention of this design is to enable the trained prompt tokens to abstractly reflect the characteristics of the matching task and reduce the overfitting of the data. In this section, we analyze the impact of different values of $k$ on performance. Figure~\ref{K_analysis_retrieval} and~\ref{K_analysis} show the performance varies with $k$ on various tasks (int Table~\ref{table:basic-datasets retrieval} and~\ref{table:basic-datasets reranking}) in retrieval and reranking stage respectively. When $k$ is $0$, the value of the prompt tokens changes with the calculation of self-attention in each layer, just like all previous prompt engineering methods. When $k$ is greater than $0$, the value of prompt tokens is fixed in $1$ to $k$ layers. Figure~\ref{K_analysis_retrieval} and~\ref{K_analysis} show that fixing the prompt tokens is always better than or equal to the previous engineering methods. Based on the experimental results on various tasks, the performance is the best when $k$ is $11$.}

\subsubsection{{Affect of Arrangement of Samples}}

{In this section, we explore the affect of the arrangement of samples in the mixed datasets consisting of multiple tasks. Specifically, we compare three arrangement strategies of samples including Random Arrangement (randomly shuffle the mixed datasets), Task Batching Alternately (all data in a batch comes from the same task and each task appears alternately in the batch), Task Batching Randomly (all data in a batch comes from the same task and each task appears randomly in the batch) and Task-balanced Batching (the data in a batch sampled balanced from each task). We report the average performance of the models trained under these three different arrangement strategies on each dataset of in-domain and out-of-domain settings. The results are shown in Figure~\ref{arrange}. In Retriever-Prompt, the loss function is in-batch contrastive loss, so ensuring that the data in a batch comes from the same task is beneficial for the model to learn the difference between positive and negative samples in the corresponding task. In Reranker-Prompt, the loss function is cross entropy, maintaining the balance between different tasks in a batch is beneficial for the model to balance each task during training and learn essential matching signals across tasks.}
\begin{table}[h]
  \caption{{Average performance of the models trained under different arrangement strategies.}}
  \label{arrange}
\renewcommand\arraystretch{1.2}
\setlength\tabcolsep{2pt}
\begin{tabular}{lcccc}
\toprule
                 & Random & Task Batching Alternately & Task Batching Randomly & Task-balanced \\ \hline
{Retriever-Prompt} & {45.15}       & {47.82} & {47.05}         & {46.70}              \\
{Reranker-Prompt}  & {63.04}       & {63.18} & {63.15}             & {64.35}  \\      
\toprule
\end{tabular}
\end{table}

\section{Related Work}
In this section, we review the previous studies on neural information retrieval methods based on text matching. {Besides, we also introduce previous work on prompt learning, it is because the core of the matching description module in our method is to obtain the description of different matching tasks to guide the learning and combination of essential matching signals. Prompt learning is the paradigm to exploit the knowledge in PLM to complete the task, and the templates constructed for the input texts can reflect the description of the task in PLM.} As for the capturing of essential matching signals, multi-task learning is an effective way to capture the signals shared between various tasks and we apply it in the essential matching module of our method. We introduce previous studies on multi-task training and prompt learning. We also introduce some recent studies on the out-of-domain generalization ability of neural information retrieval models.

\subsection{Neural Information Retrieval Based on Text Matching}
In this subsection, we will review some related studies on two stages in the neural information retrieval pipeline (i.e. retrieval and reranking) from the perspective of three essential matching signals including exact matching, semantic matching, and inference matching.
\subsubsection{Neural Retrieval Based on Text Matching}
Neural retrieval is the first stage in the information retrieval system that efficiently and accurately obtains candidate subsets from the massive document base. In exact matching, DrQA~\cite{open-domain-qa} combines TF-IDF weighted bag-of-words vectors and bigram to represent the text. This method shows good performance in open-domain QA. In semantic matching, models based on single semantic text representation such as DSSM~\cite{dssm}, CDSSM~\cite{cdssm}, and ARC-I~\cite{arc1} represent text as a dense vector to get the matching score. These models ensure retrieval efficiency and are often used in DR and QA. More recently, PLMs have achieved state-of-the-art neural retrieval performance and become the mainstream neural retrieval method that can be used to represent different matching signals based on the training on in-domain datasets. DPR~\cite{dpr} uses BERT to encode query and passage into high-dimensional vectors and train the model with contrastive learning to obtain a dense vector retriever. ANCE~\cite{ance} reveals the effect of hard negatives in the training of dense retriever and propose a method to refresh the corpus index during training and retrieve the hard negatives for queries in the training set. RocketQA also optimizes the utilization of samples during training by dense retrieval from the three levels including expanding batch size, data enhancement, and denoising negative sampling. Some methods such as Condenser~\cite{condenser} propose to use the unsupervised pre-training to further enhance the representation ability of the PLM for text. ARR~\cite{ARR} uses the reranker as a discriminator and exploits adversarial training to improve the ability of dense retrieval. TAS-B~\cite{tas-b} distills knowledge from reranker to dense retrieval and boosts performance.

\subsubsection{Neural Reranking Based on Text Matching}
Reranking is the finer ranking stage for the candidate subsets retrieved from the first stage, which models the interaction between texts more complexly. In exact matching, DRMM~\cite{drmm} considers query item importance and is suitable for DR. In semantic matching, interaction-based text matching models such as DeepMatch~\cite{deepmatch}, ARC-\uppercase\expandafter{\romannumeral2}~\cite{arc2}, MatchPyramid~\cite{matchpyramid}, ESIM~\cite{esim} and Match-SRNN~\cite{SRNN} can describe the fine-grained semantic matching relationship between two texts. They are often used in short text matching tasks such as PI, NLI, and RD. In inference matching that needs models to infer new information from the text, asymmetric neural text matching models such as DeepRank~\cite{deeprank} and RE2~\cite{re2} are suitable. Recently, PLM-based framework such as Monobert~\cite{monobert} that concatenates two texts and uses the self-attention mechanism for deep interaction achieves state-of-the-art performance in reranking.

The previous neural information retrieval models are only suitable for one of the specific tasks according to their specifically designed structures and mechanisms. Even though PLMs can be applied in multiple text matching tasks, only task-specific models that are fine-tuned on specific tasks can achieve good performance. When using the traditional fine-tuning method to train PLMs on mixed datasets of multiple matching tasks, the performance of the model drops seriously~\cite{linguistic}. Our approach focuses on capturing the essential information that can be used across tasks and domains and adapting them to the different tasks and domains to improve the generalization ability of NIR models.

\subsection{Multi-Task Learning for Natural Language Processing Tasks}
{In this section, we review multi-task learning in natural language processing tasks from the perspective of traditional multi-task learning and multi-task prompt learning.}
\subsubsection{Traditional Multi-Task Learning}
{In order to enable a single model to handle multiple tasks effectively, a common strategy is to train the model on mixed datasets that encompass various tasks. One straightforward approach is to directly combine the datasets of each task without incorporating any task-specific marks, and then fine-tune pre-trained language models (PLMs) on these mixed datasets. For example, Alon et al.\cite{multiqa} propose a method called MultiQA. They train models on multiple datasets without task-specific tokens and find that it leads to robust generalization in reading comprehension tasks. Another approach, introduced by Jean et al.\cite{multi-retrieval}, is a multi-task retrieval model designed for knowledge-intensive tasks. They utilize the mixed datasets provided by KILT~\cite{KILT} without using any task-specific tokens. Some approaches include additional task-specific components that contain parameters tailored to each task. Liu et al.\cite{mt-dnn} propose MT-DNN, which employs a shared transformer encoder to encode multiple natural language understanding (NLU) tasks and incorporates task-specific layers for each task. Additionally, there are methods that transform multiple tasks into a unified question-answering format, such as MQAN\cite{MQAN}, UnifiedQA~\cite{unifiedqa}, and T5~\cite{t5}. Li et al.\cite{ie} apply this method to information extraction, while Chai et al.\cite{tc} and Puri et al.~\cite{zero-shot-tc} use it in text classification.}

{The main focus of the above methods is not information retrieval. Furthermore, some of these methods fail to fully utilize differentiating marks, while others introduce additional task-specific layers or require extremely large-scale models. In comparison, NIR-Prompt aims to leverage multi-task training to capture essential signals that can be applied across tasks and domains, and combine these signals to adapt to different tasks and domains, thereby improving the multi-task generalization capability of PLMs in neural information retrieval. Notably, when using the NIR-Prompt model for prediction, there is no need to add any task-specific layers.}

\subsubsection{{Multi-task Prompt Learning}}
{In recent times, there has been an increasing trend in utilizing parameter-efficient methods for multi-task learning and transfer learning across various works.~\cite{explore_transfer} conduct an extensive study of the transferability across various NLP tasks.~\cite{transfer} investigates the transferability of prompts across different tasks. PANDA~\cite{panda} proposes a metric to accurately predict prompt transferability and a novel method to transfer the knowledge from source to target prompt via knowledge distillation. AdapterFusion~\cite{adapterfusion} proposes a two-stage learning algorithm that leverages knowledge from multiple tasks. HyperFormer~\cite{hyperformer} learns adapter parameters for all layers and tasks by generating them using shared hypernetworks. ATTEMPT~\cite{attempt} exploits attentional mixtures of soft prompts for parameter-efficient multi-task learning.} 

\subsection{Prompt Engineering}
{Prompt learning is an innovative tuning strategy for pre-trained language models (PLMs) that converts multiple natural language processing (NLP) tasks into [MASK] prediction format using language models. This is achieved by incorporating a template into the input texts. The creation of an effective template is crucial for successful prompt learning. Generally, there are two main categories for template creation: manual template engineering and automated template learning~\cite{promptzongshu}.}

{In manual template engineering, Petroni et al.\cite{lama} design templates to extract knowledge from PLMs. Schick et al.\cite{pet} propose Pattern-Exploiting Training (PET), which combines few-shot learning with templates and transforms specific tasks into cloze tasks. Although manual template creation can address many tasks, it has certain limitations: 1) it requires extensive knowledge and expertise to design templates, and 2) manually created templates may not always be globally optimal~\cite{prefix}. To overcome these challenges, several automated template learning methods have been proposed, such as token-based gradient searching~\cite{searching}, mining training corpus~\cite{mining}, Prefix Tuning~\cite{prefix}, and P-tuning~\cite{p-tunning}. Recently, there have been studies focusing on pre-training prompts on multiple tasks, such as SPoT~\cite{spot}, PPT~\cite{ppt}, ExT5~\cite{ext5}, FLAN~\cite{instruction-learning}, ZeroPrompt~\cite{zeroprompt}, and zero-shot for task generation~\cite{zero-generation}.} {Different from previous research, our work does not primarily aim to improve the performance of prompt learning. Instead, our focus is on utilizing prompt learning to enhance the generalization capability of neural information retrieval models. Specifically, we employ prompt learning to separate the process of capturing signals and combining them in text matching. The key role of prompt learning in our method lies in obtaining task descriptions for various matching tasks within PLMs and utilizing these descriptions to combine essential matching signals for different tasks and domains. Additionally, we explore the relationships among task prompt tokens and discover that new matching task prompts can be constructed by linearly combining other learned task prompt tokens.}

\subsection{Out-of-domain Generalization of Neural Information Retrieval}
The generalization of neural information retrieval models across domains (i.e. out-of-domain multi-task performance of our research questions) has recently received attention from researchers. In general, the out-of-domain generalization of neural information retrieval is poor, even weaker than the traditional word-overlap-based method such as BM25~\cite{bm25}. A thorough examination of the generalization of dense retrieval is performed~\cite{Examination} and discusses the key factors that affect the performance of zero-shot dense retrieval including the vocabulary overlap, query type distribution, and data scale. BEIR~\cite{beir}, the heterogeneous benchmark for testing the generalization ability of retrieval models is designed to promote the relevant research. MoDIR~\cite{MIDR} introduces a momentum method to learn the domain-invariant by adversarial training on the source and target domain. DDR~\cite{ddp} disentangles the retrieval model to Relevance Estimation Module (REM) for modeling domain-invariant matching patterns and several Domain Adaption Modules (DAMs) for modeling domain-specific features of multiple target corpora to propose an adaptable dense retrieval framework. However, both of these methods require the data from the target domain to be acquired in training. They are unsupervised domain adaptation, not generalization, which is inconsistent with our settings (no target domain data required).

\section{Conclusion}
In this paper, we point out that although there are some differences among the various information retrieval tasks, there are still essential matching signals shared by the various tasks, such as exact matching, semantic matching, and inference matching. If the model can capture and exploit these signals, the generalization ability of the model across tasks and domains will be improved. With this intuition, we propose a neural information retrieval training framework called NIR-Prompt consisting of Essential Matching Module (EMM) and Matching Description Module (MDM) based on the idea of decoupling the process of signal capturing and signal combination. MDM uses the method of prompt learning to obtain the description of different tasks in the pre-trained language model. EMM is trained on diverse mixed datasets and combined with the guidance from the task descriptions in MDM to capture essential matching signals and adapt these signals to different tasks. Based on this, a generalized neural information retrieval pipeline consisting of retrieval and reranking is constructed. The experimental results on eighteen public datasets and a heterogeneous benchmark for testing the generalization ability of retrieval models show that our method yields better in-domain multi-task, out-of-domain multi-task, and new task adaptation performance for dense retrieval, reranking, and the entire neural information retrieval pipeline compared to the traditional fine-tuning paradigm.

\bibliographystyle{ACM-Reference-Format}
\bibliography{my}










\end{document}